\documentclass[aip,twocolumn,reprint]{revtex4-1}

\usepackage{graphicx}
\usepackage{bm}
\usepackage{color}
\graphicspath{{./figures/}}

\begin{document}

\title{Encoding Mechano-Memories in Actin Networks}

\author{Sayantan Majumdar}
\affiliation{James Franck Institute, The University of Chicago, Chicago, Illinois 60637, USA.}
\author{Louis C. Foucard}
\affiliation{Department of Chemistry \& Biochemistry, UCLA, Los Angeles, USA}
\author{Alex J. Levine}
\affiliation{Department of Chemistry \& Biochemistry, UCLA, Los Angeles, USA}
\affiliation{Department of Physics \& Astronomy UCLA, Los Angeles, USA}
\affiliation{Department of Biomathematics UCLA, Los Angeles, USA}
\author{Margaret L. Gardel}
\affiliation{James Franck Institute, The University of Chicago, Chicago, Illinois 60637, USA.}

\date{\today}

\begin{abstract}
Understanding the response of complex materials to external force is central to fields ranging from materials science to biology. Here, we describe a novel type of mechanical adaptation in cross-linked networks of F-actin, a ubuiquitous protein found in eukaryotic cells. We show that shear stress changes its nonlinear mechanical response even 
long after that stress is removed. The duration, magnitude and direction of forcing history all impact changes in mechanical response. The `memory' of the forcing history is long-lived, but can be erased by force application in the opposite direction.  We further show that the observed mechanical adaptation is consistent with stress-dependent changes in the nematic order of the constituent filaments. Thus, this mechano-memory is a type of nonlinear hysteretic response in which an applied, ``training'' strain modifies the nonlinear elasticity.  This demonstrates that F-actin networks can encode analog read-write mechano-memories, which can be used for adaptation to mechanical stimuli.
\end{abstract}


\maketitle

\begin{figure}
	\centering
	\includegraphics [width = 8 cm]{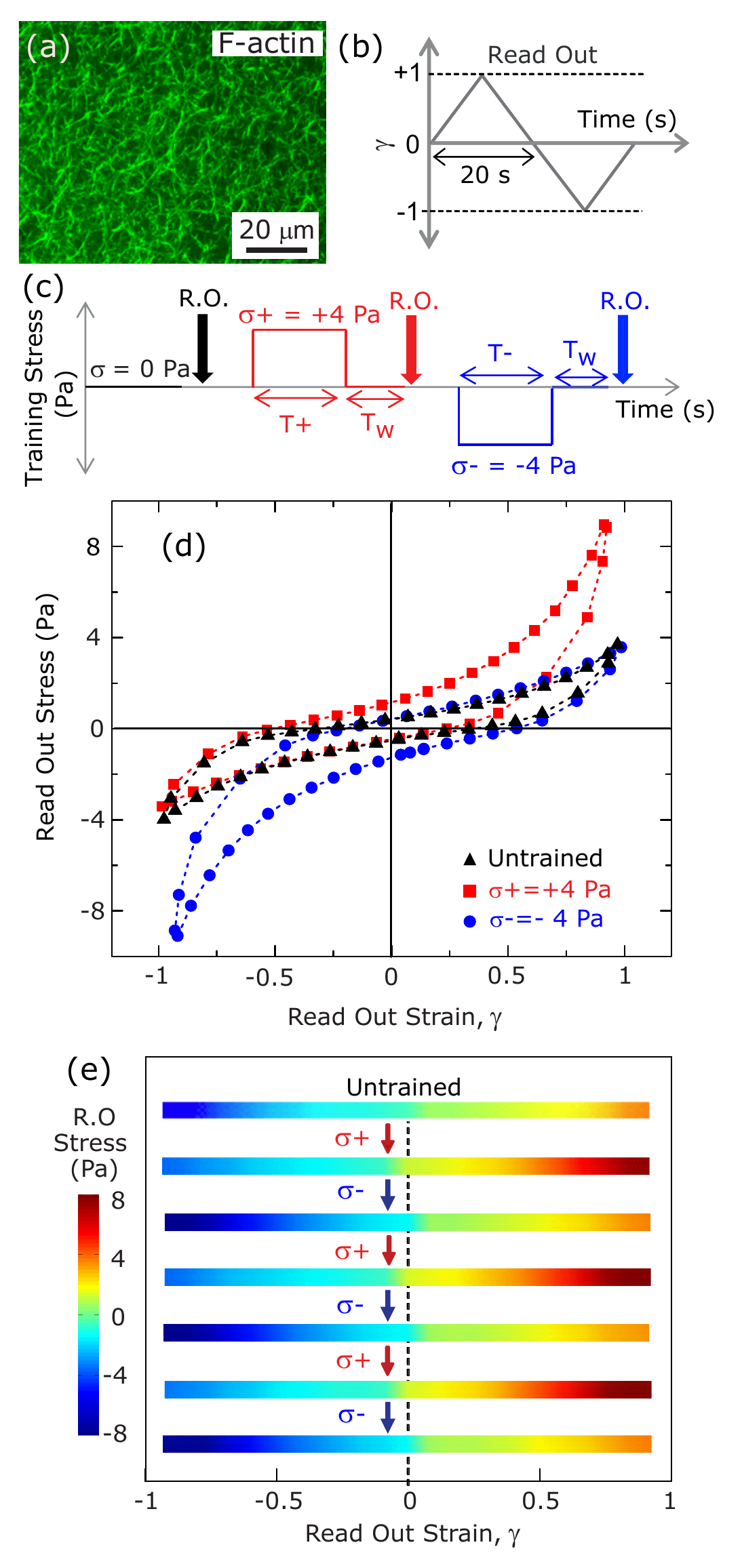}
	\caption{\textbf{History dependent shear response in cross-linked actin network.} \textbf{a}, Typical confocal image plane of a fully polymerized 3-D actin filament network cross-linked with Filamin A. The concentration of actin ($c_a$) is 24 $\mu M$ and the molar ratio of cross-linker to actin is 5\%. \textbf{b}, Readout protocol under controlled strain at a constant shear rate. \textbf{c}, Training and readout sequence for probing reversible mechano-memory of a freshly polymerized sample. \textbf{d}, Lissajous plots (stress vs strain) generated from the readouts described above. The training and readout sequences are applied as in (\textbf{c}) with training time $T_{\pm}$ = 250 s. \textbf{e}, Stress vs strain represented by heat maps obtained from the readouts between applied training stresses of magnitude 4 Pa applied alternatively along positive and negative directions, for 250 s.}
	\label{F1}
\end{figure}

Biological cells sense mechanical stimuli and use these cues to control their physiology \cite{sperelakis2012cell}. Elucidating such mechanosensory mechanisms in living matter has implications for both cell biology and smart materials design \cite{cui2015dynamic}.  While molecular-scale mechanotransduction mechanisms are well appreciated \cite{iskratsch2014appreciating, wang1993mechanotransduction}, how these are integrated to control mechanical response at cellular length scales is not well understood.  Networks of the semi-flexible biopolymer actin are essential determinants of the mechanical behaviors of eukaryotic cells \cite{blanchoin2014actin}. Previous work has shown that the mechanical response of biopolymer networks can be regulated either transiently \cite{gardel2004elastic} or irreversibly \cite{schmoller2010cyclic} by external force, necessitating mechanochemical feedbacks to enable long-lived and dynamic mechanical adaptation \cite{bieling2016force}.  
\newline
\newline
When actin is polymerized in the presence of cross-linking proteins, space-spanning networks are formed (Fig.\ref{F1}a). The mechanical response of actin networks is viscoelastic \cite{tharmann2007viscoelasticity, muller2014rheology,foucard2012a} and highly nonlinear \cite{storm2005nonlinear}. Typically, the strain stiffening of actin networks is 
reversible with no dependence on the history of applied stress \cite{gardel2004elastic}. However, polymeric systems of both biological and synthetic origin can also demonstrate irreversible work hardening \cite{schmoller2010cyclic} and other permanent rheological changes in response to applied stress \cite{diani2009review, schmoller2013similar, munster2013strain}. The irreversibility of work hardening limits the potential for these materials to dynamically modify their mechanical properties or sense external force. 
\newline
\newline
Here, we probe the rheology of a physiologically relevant network of actin filaments cross-linked with the protein filamin (FLN). FLN-crosslinked actin networks exhibit a dramatic strain stiffening \cite{gardel2006prestressed, kasza2007cell, gardel2008mechanical}. FLN cross-links are also highly dynamic, with an off rate of $\sim 1$ s$^{-1}$ ~\cite{goldmann1993analysis} that results in a broad spectrum of stress relaxation \cite{broedersz2010cross}. We polymerize 24 $\mu$M G-actin in the presence of 5\% FLN within a parallel plate rheometer and probe the 
mechanical properties via a symmetric, triangular strain `read out' (RO) pulse (Fig.\ref{F1}b). At low strain, a linear stress-strain relationship is observed but the stress increases 
nonlinearly at higher strains. The concavity in the stress-strain curve reflects the nonlinear strain stiffening; this behavior is symmetric for strains in the opposing direction 
due to the isotropic nature of the network (black triangles, Fig.\ref{F1}d). To explore how applied stress can alter this rheology, we apply a large constant stress in the 
positive direction ($\sigma_{+}$ = 4 Pa) for 250 s (Fig.\ref{F1}c).  Upon removal of the stress, we wait for 50 s and read out the mechanical response by applying a 
RO pulse.  Surprisingly, we observe a dramatic increase in the stress in the direction of the training with nominal reduction in stress in the opposite direction 
(Fig.\ref{F1}d, red squares); this indicates that the mechanical response is altered by the prior stress application. To explore the reversibility of this effect, 
we apply a constant stress of 4 Pa in the opposite direction ($\sigma_{-}$) for 250 s. Remarkably, after the negative training pulse, an enhanced stiffening is observed at 
negative strains (Fig.\ref{F1}d, blue circles) and all evidence of the previous positive training pulse is erased. To underscore the reversibility of this training, we apply 
successive cycles of positive and negative training. In Fig.\ref{F1}e, the stress measured between each cycle is plotted as a function of strain over six training cycles. 
These data demonstrate the ability to form and erase mechanical memories with robust reversibility. Such `mechano-memory' is a type of nonlinear hysteretic 
response in which applied `training' strain modifies the nonlinear elastic response of the material to subsequent loading.
\begin{figure*}
	\centering
	\includegraphics[width=12 cm]{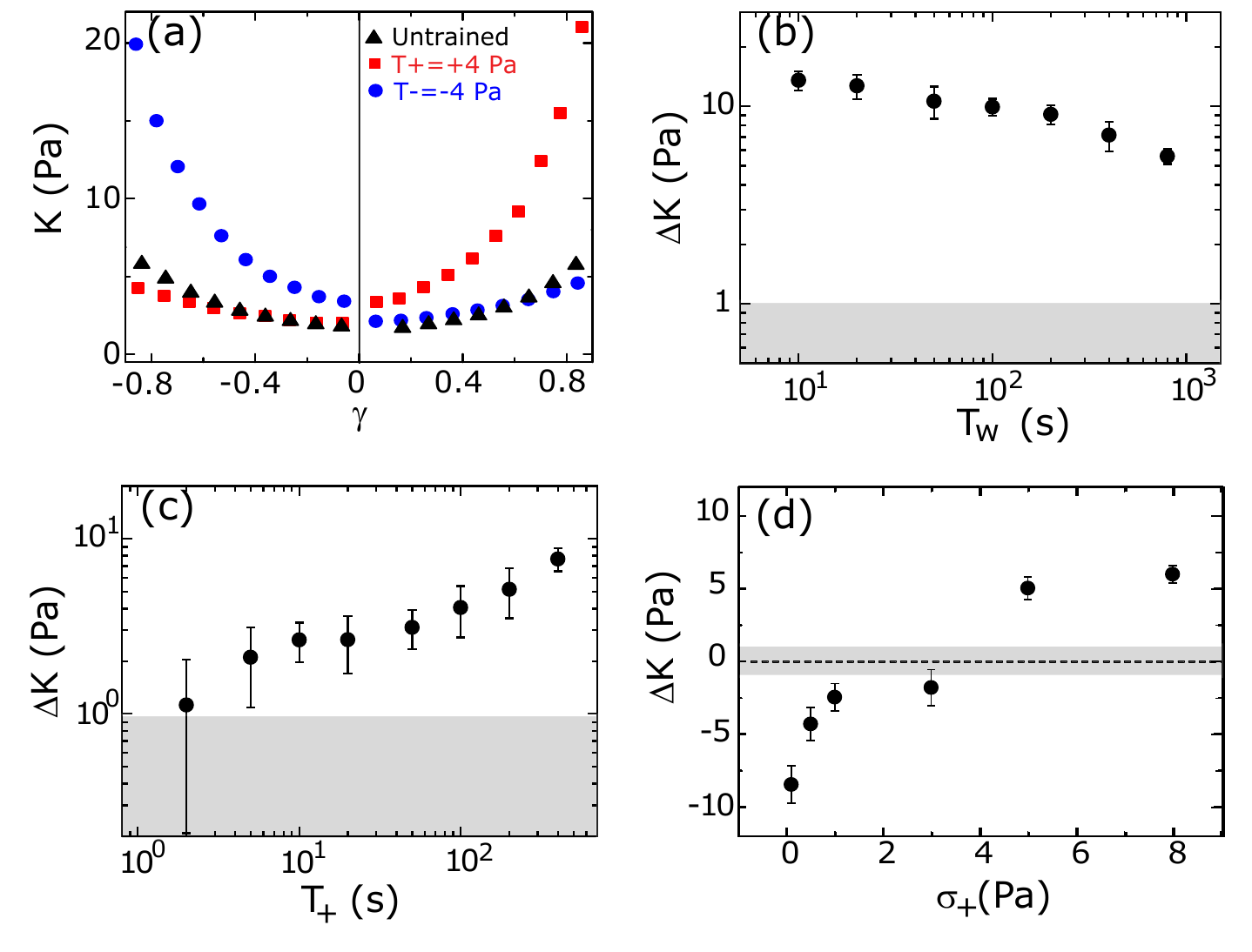}
	\caption{\textbf{Properties of mechano-memory: reversibility, encoding and rewriting.} \textbf{a}, Shear modulus ($K$) as a function of strain ($\gamma$) obtained from Fig.\ref{F1}d for an untrained, positively trained and negatively trained sample as indicated. \textbf{b}, Decay of mechano-memory (originally formed with training stress $\sigma_+$ = 4 Pa applied for a training time of $T_+$ = 250 s) quantified by the magnitude of $\Delta K$ estimated at $|\gamma|$ = 0.4 as a function of waiting time. \textbf{c}, Building up of mechano-memory ($\sigma_+$ = 4 Pa) quantified by the magnitude of $\Delta K$ at $|\gamma|$ = 0.4 as a function of training time. In both \textbf{b} and \textbf{c}, the shaded region indicates the baseline asymmetry in an untrained sample. \textbf{d}, Erasing and rewriting mechano-memory for a sample initially trained in the negative direction by $\sigma_{-}$ = 5 Pa and $T_{-}$ = 250 s; $\Delta K$ estimated at $|\gamma|$ = 0.4 is plotted when an increasing magnitude of $\sigma_{+}$ is applied on this sample for $T_{+}$ = 100 s. In \textbf{b}, \textbf{c} and \textbf{d} the error bars are estimated from the average standard deviation of $K$ vs $\gamma$ with respect to a mean curve drawn through them (see Methods). }
	\label{F2}
\end{figure*}

\begin{figure*}
	\centering
	\includegraphics[width=16 cm]{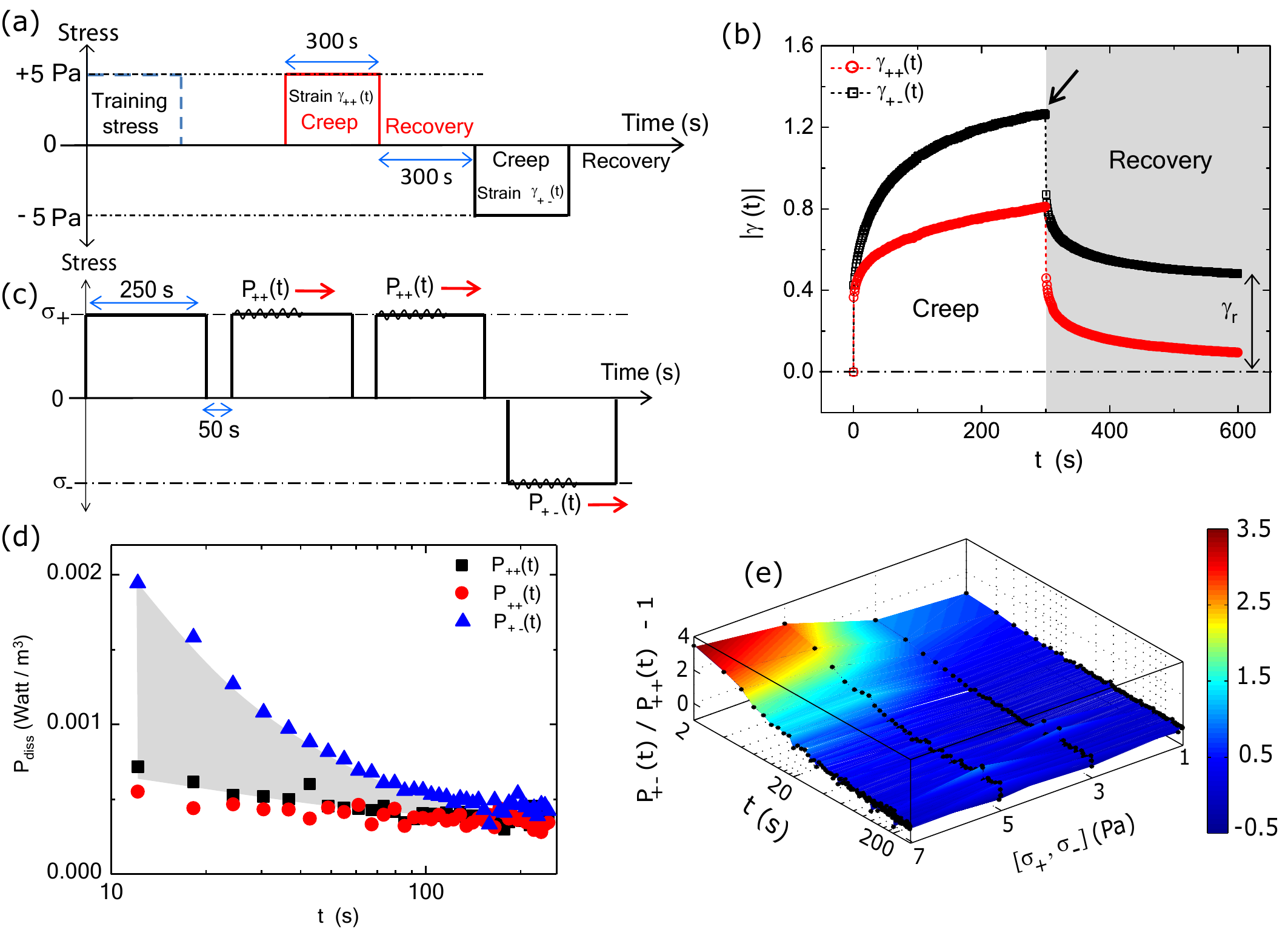}
	\caption{\textbf{History dependent creep behavior and energy dissipation} \textbf{a}, Schematic of the protocol to measure history dependent creep response: An untrained sample ($c_a$ = 24 $\mu$M actin + 5 \% FLN) is first trained by $\sigma_+$ = 5 Pa stress in the positive direction for 300 s. After waiting for 300 s, a stress of magnitude 5 Pa is again is applied for 300 s either in the training direction (positive) or opposite to it (negative) and then the stress is switched off ($\sigma$ = 0 Pa) for 300 s. \textbf{b}, Temporal evolution of the magnitude of the creep strain when the stress pulse is applied in the training direction $|\gamma_{++}(t)|$ or opposite $|\gamma_{+-}(t)|$ to it. The residual strain ($\gamma_r$) estimated after 300 s of recovery is indicated for the case of strain evolution $\gamma_{+-}(t)$. \textbf{c}, Schematic of the protocol to estimate energy dissipation in rewriting a memory. \textbf{d}, An untrained sample ($c_a$ = 24 $\mu$M actin + 5 \% FLN) is first trained by $\sigma_+$ = 4 Pa stress in the positive direction for 250 s. Time dependent power dissipation is estimated (see Method) by superposing a small A.C. component of stress (amplitude 0.4 Pa) on the D.C. stress pulses (magnitude 4 Pa) applied in the same ($P_{++}(t)$) or opposite direction ($P_{+-}(t)$) w.r.t the initial training direction. The integrated area between $P_{++}(t)$ and $P_{+-}(t)$ indicated by the gray colored region gives the additional dissipated energy in re-writing a memory. \textbf{e} A sample is first positively trained by $\sigma_{+}$ applied for 250 s. We then estimate the time dependent power dissipation ($P_{+-}$) when a stress $\sigma_-$ ($|\sigma_{+}| = |\sigma_{-}|$) is applied in the negative direction on this sample. We normalize $P_{+-}$ by the power dissipation ($P_{++}$) when $\sigma_{+}$ is applied on the initially positively trained sample. Each curve (solid circles) is obtained for one pair of $\sigma_{+}$ and $\sigma_{-}$ values. The 3-D plot is generated by varying the magnitude of such pair [$\sigma_{+}$, $\sigma_{-}$].}
\label{F3}
\end{figure*}

\begin{figure*}
	\centering
	\includegraphics [width=14 cm]{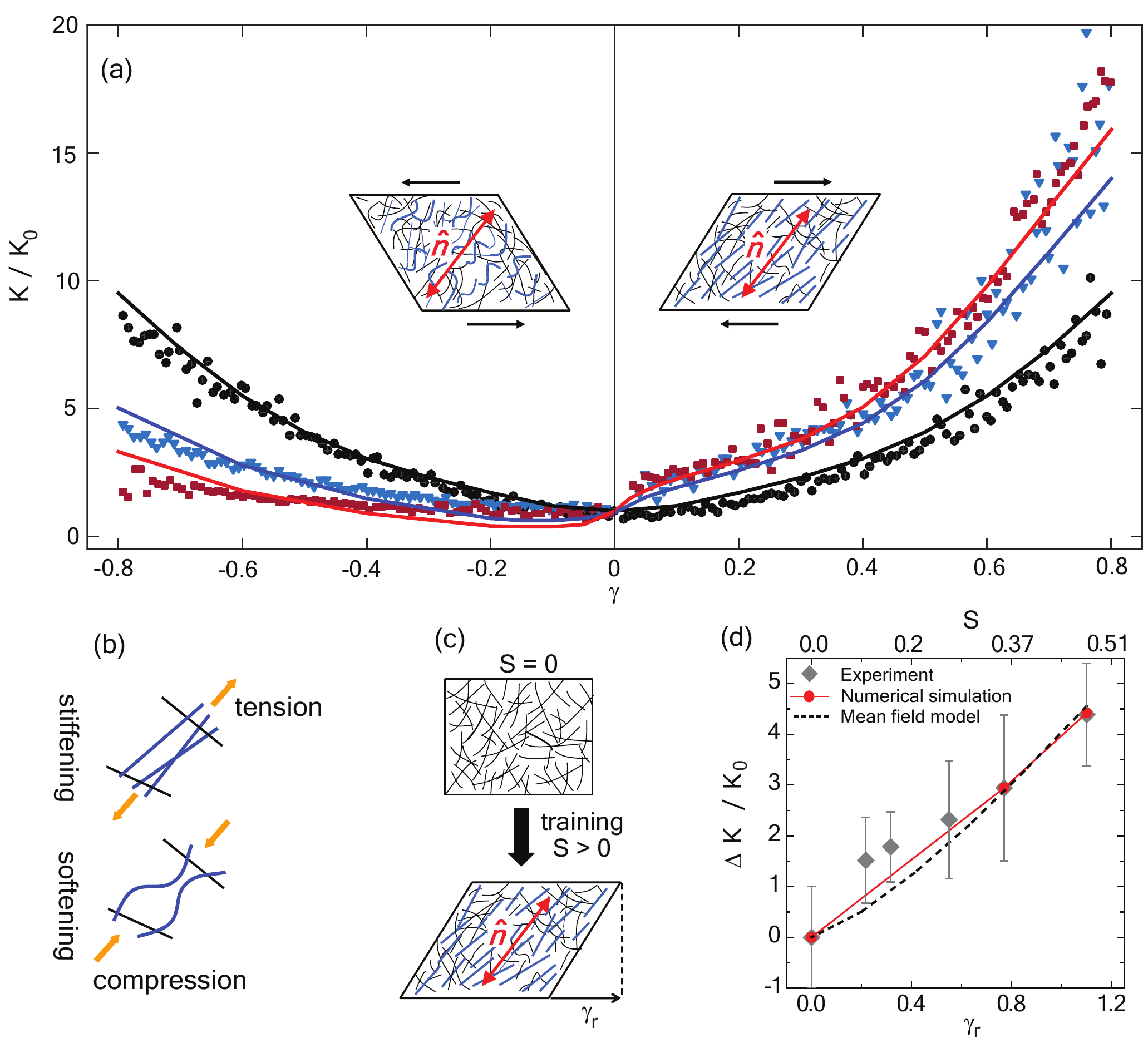}
	\caption{\textbf{The nematic order encodes the history dependent shear response of actin networks.} \textbf{a}, Measurement and prediction of the relative stiffening as a function of  shear strain $\gamma$, for different training times $T_+$ as described in the text. The training stress $\sigma_+$ = 4 Pa in all cases. Full markers represent experimental data while continuous curves the corresponding predicted data from simulations. \textbf{b}, Cartoon depicting that for a positively trained sample, positive RO strains put most filaments are under tension which lead to a larger overall network stiffness. Conversely, negative strains put most filaments under compression, the network enters a soft bending mode and the overall stiffness decreases. \textbf{c}, The residual strain measured experimentally after training is used to compute an equivalent nematic order S under the assumption of affine deformation. This nonlinear behavior is magnified as the nematic order increases. \textbf{d}, We observe an excellent agreement between the relative stiffening measured experimentally (discrete points) at $|\gamma|$=0.4 for different values of residual strains (e.g. different training time) and predicted by the simulations (red line with symbol) for the corresponding nematic order, as well the prediction from the simple mean field model (black dashed line). } 
	\label{F4}
\end{figure*}

The nonlinear mechanical response can be characterized by plotting the differential shear modulus, $K = \frac{d\sigma}{d\gamma}$, obtained by differentiating the Lissajous curves numerically along the increasing strain magnitude, as a function of strain $\gamma$. For an untrained sample, $K (\gamma)$ is symmetric around $\gamma$ = 0 and shows stiffening observed at high strains in both the positive and negative direction (black triangles, Fig.\ref{F2}a). For positively or negatively trained samples, the symmetry in $K$ as a function of $\gamma$ is clearly broken: $K$ is enhanced for $\gamma >$0 for a positively trained sample (red squares in Fig.\ref{F2}a), whereas, for a negatively trained sample the enhancement happens for $\gamma <$0 (blue circles in Fig.\ref{F2}a). The parameter $\Delta K = K_+ - K_-$, where $K_+$ and $K_-$ are the moduli measured at strain magnitude $0.4$ in the positive and negative directions, respectively, is used to quantify the strength of the memory formed (Methods). 
Due to the symmetry in the untrained sample, assignment of the direction of positive strain and stress is arbitrary. Once chosen, the sign convention is used consistently in all subsequent measurements for that sample.

To query whether the mechanomemory is long-lived, we vary the time ($T_w$, Fig.\ref{F1}c) between removal of the training stress and the read out measurement.  
We find that, $\Delta K$ decreases weakly up to $T_w$ of nearly 1000 seconds (Fig.\ref{F2}b), indicating that the training pulses encode a semi-permanent memory into the network. 
To examine how the training parameters can be used to modify the strength of the mechanomemory, we vary the magnitude and duration of the training stress. The duration 
of training tunes $\Delta K$, with $\Delta K$ monotonically increasing from 2 to 8 Pa as training time $T_+$ is increased from 10 to 400 s (Fig.\ref{F2}c). Likewise, 
the training stress magnitude also tunes $\Delta K$ (see Fig.S1).  For a sample maximally trained in the negative direction ($\sigma_-$ = 5 Pa, $T_-$ = 250 s), $\Delta K \approx$ -8 Pa.  When a positive training pulse $\sigma_{+}$ = 0.5 Pa is applied for $T_+$ = 100 s, $\Delta K$ increases to $\approx$ -5 Pa. When $\sigma_{+}> 4$ Pa, $\Delta K >$ 0 (Fig.\ref{F2}d). 
After maximal training, it is possible to recover a mechanical response similar to an untrained sample with a suitable stress pulse in the opposing direction (Fig.S1).

To explore the effects of training stress, we study the time dependent changes in strain (i.e., creep) and recovery behavior.  First, a training stress in the positive 
direction is applied for 300 s (Fig.\ref{F3}a).  For a subsequent stress pulse in the same direction ($\sigma_+$ = 5 Pa, $T_+$ = 300 s), the resulting strain magnitude 
$\gamma_{++}$ increases to 0.8.  After the stress is removed, the residual strain at 300 s, or $\gamma_r$, is approximately 0 (Fig.\ref{F3}b).  By contrast, when an identical 
stress pulse is applied in the direction opposite to the direction of training, the strain magnitude $|\gamma_{+-}|$ increases up to 1.2. After the stress is removed, $\gamma_r$ = 0.4 (Fig.\ref{F3}b). This increased residual strain reflects a plastic-like shape deformation upon stress reversal.  

The differences in the creep observed in response to different directions of applied stress indicate that the system is effectively more elastic in the training direction compared to the opposite direction. Another way to think about this is that there is more energy dissipated in a creep measurement that occurs after reversal of stress direction. We directly calculate the dissipated power from the creep measurements and observe an an enhanced dissipation when the the direction of stress is changed (Fig.S2).  The area under the power dissipation curves for these two different loading directions gives an estimate of additional energy that gets dissipated in reorganizing the system. We observe an increase in energy dissipation as a function of applied stress (Fig.S2). Due to the noise in calculating energy dissipation directly from creep measurements, we use an alternate strategy (Methods) of superimposing a small, oscillatory stress during the creep measurement to probe power dissipation (Fig.\ref{F3}c).  Using this alternate strategy, we also measure an enhanced dissipation when the system is loaded in the direction opposite to that of training direction (Fig.\ref{F3}d). We then calculate the dissipative power required to rewrite a memory of similar strength but different sign (i.e., from +$\Delta K$ to -$\Delta K$) and normalize it by that dissipated in the absence of a new memory formation ($P_{+-} / P_{++}$) (Methods). We find that the power dissipation increases dramatically for stresses $>$ 3 Pa and mostly contained within the first 100 s of stress application (Fig.\ref{F3}e). This underscores the stress- and time-dependent work done by the external mechanical stresses to dynamically alter the network mechanical response.
\newline
\newline
To elucidate the mechanism of mechanomemory, we compare the  normalized differential modulus as a function of strain (Fig.\ref{F4}a) for networks trained ($\sigma_{+}$ = 4 Pa) for a duration of  $T_+$ = 0 s (untrained, black circles), $T_+$  = 200 s (upside down triangles) and $T_+$  = 400 s (squares)  with a simple two-dimensional finite-element simulation (lines). In the simulation 2 $\mu$m long stiff filaments are cross linked by 100 nm long, flexible linkers to represent FLN (see Fig.S3). For the isotropic networks, the strain dependencies in the differential modulus arises primarily from non-linearities in the constitutive elements (e.g. cross-linkers and filaments) \cite{Marko1995}.
\newline
\newline
Previous work has shown that small degree of filament alignment, or nematic order, leads to a dramatic change in the nonlinear elastic response \cite{Foucard2015, Missel2010}. Specifically, when a network of rigidly cross linked semiflexible filaments is sheared so that the compression direction lies along the nematic director, softening is observed at strains of less than $0.1 \%$ due to the buckling of the filaments under compression (Fig.\ref{F4}b, bottom schematic). When the same network is sheared in the opposite direction so that the same aligned filaments are now tensed, the system exhibits strain stiffening (Fig.\ref{F4}b, top schematic). These nonlinear elastic effects make the stiffness versus strain curve highly asymmetrical. 
\newline
\newline
This \textit{collective} nonlinearity, i.e. nonlinearity emerging from the interaction of many filament, is the dominant effect for small and moderate strains \cite{Foucard2015} (Fig.\ref{F4}a). At larger strains, the \textit{constitutive} nonlinear stiffening of individual cross-linkers contributes as well \cite{Marko1995} (see Fig.S4). This leads to a strong strain stiffening of the network at larger values of $\gamma$, where the collective nonlinearity combines with the constitutive nonlinearity to produce additional asymmetric strain stiffening.  
\newline
\newline
In order to compare our model to the experiment, we estimate the degree of nematic order produced by training. Here, we do not consider in the simulation the training and cross-linking dynamics. Rather, we rely on the observed post-training residual strain $\gamma_{r}$ (Fig.\ref{F3}b, also see Fig.S5), which provides a direct measure of the change in the elastic reference state. Presuming an affine deformation, we then calculate the expected nematic order induced by filament reorientation by this shear (see Method). Using this simple, geometric model to infer the degree of nematic order $S(\gamma_{r})$, we numerically construct an equivalent semiflexible network for each observed $\gamma_{r}$ and determine $\Delta K(\gamma_{r})$ computationally.
\newline
\newline
The tangent modulus $K(\gamma)$, normalized with respect to its value before training is shown for one untrained (black) network and two networks trained for 200s (blue) 
and 400s (red) as lines in Fig.\ref{F4}a. We find excellent agreement with the experimental data.  We then plot $\Delta K$ as a function of residual strain (Fig.\ref{F4}d). The agreement between the observed and simulated mechanomemory development strongly suggests that the development of nematic order underlies the observed changes in elastic nonlinearities. Interestingly, we see evidence of such nematic order in the network under shear deformations (Fig.S6), which is qualitatively consistent with our model assumptions.

To further investigate the role of nematic order in the network and consequently the emergence of buckling in its nonlinear response, we build a mean field network model composed of linear elastic filaments with frozen nematic order $S$, and incompliant cross-linkers. We account for filament buckling simply by neglecting the mechanical contribution from filaments under compressive loads exceeding their Euler criterion~\cite{Timoshenko1986} (see Supplement). Figure~\ref{F4}d plots $\Delta K$ predicted by the mean-field model (dashed black line) agrees with both the simulation and experiment. Based on the mean field model, we suggest that proposed dependence of filament buckling on the degree nematic alignment in the gel serves as a minimal model of mechanomemory effects.

Our study demonstrates the reversible formation of mechanomemories in physiologically relevant cross-linked actin networks.
Previous studies of strain-history dependence in the rheology of biopolymer networks observed {\em irreversible} effects attributed to permanent changes in network bundling \cite{schmoller2010cyclic, munster2013strain, schmoller2013similar}. Importantly, using the identical cyclic hardening protocol used in \cite{schmoller2010cyclic}, we do not observe work hardening (Fig.S7).  
This confirms the effect we describe here is distinct from irreversible work hardening observed in previous studies.  Future work is needed to understand how training stress and 
duration as well as network composition can control the formation of either reversible or irreversible stress-dependent changes in the mechanical response. 

The field-dependent hysteresis in mechanical response we describe is highly reminiscent of other hysteresis phenomena (e.g. magnetism). For instance, our data suggests that the stress-dependent changes in mechanical response arise from the formation of nonequilibrium, but long-lived domains of partial nematic order in the network that strongly modifies the nonlinear response of the network to strain. Care must be taken to distinguish such hysteresis from the ubiquitous rate-dependent hysteresis observed in viscoelastic materials. Our study demonstrates that that the stress-dependent changes in mechanical response can be understood solely in terms of the development of long-lived nematic order.

Previous work on nematic elastomers \cite{pujolle2001observation, kannan1993rheology} and worm-like micelles \cite{kadoma1998flow} has demonstrated the importance of 
nematic order on mechanical response.  Here, the ability to rewrite and erase memories comes from the fact that the lifetime of the nematic order is highly stress dependent. 
Networks of elastically nonlinear elements (e.g., filaments subject to softening under compression in response to buckling) that can structurally reorganize under load (develop nematic order) point to an avenue for the development of smart adaptive materials based on these cytoskeletal structures.
There remain open questions regarding how mechanomemory depends on network density and composition.  We have focused here on actin networks 
cross-linked with the flexible and dynamic cross-linker filamin.  We observe qualitatively similar, but quantitatively distinct, behavior using the cross-linking protein 
$\alpha$-Actinin which is more rigid and shorter, but with similar binding kinetics (Fig.S8).  This suggests that cross-linker properties may prove critical to 
controlling the details of the stress-dependent changes in nematic order. Moreover, understanding the role of network density and, specifically 
on the transition from non-affine to affine deformations \cite{Foucard2015, head2003deformation, wilhelm2003elasticity} is 
also likely to be important. Exploring these properties in biopolymer networks will enable a more complete understanding of the 
control parameters to inform design of materials constructed from synthetic analogs.

Our study demonstrates the possibility to reversibly alter the mechanical response of network to enable one to write, read and erase mechanomemories.  In fact, this system may have the potential to encode multiple memories \cite{paulsen2014multiple} through the formation of multiple nematic domains with differing nematic directors, leading perhaps to new classes materials with a programmable and complex nonlinear elastic response. Finally, our results also suggest a mechanism of observed force adaptation in living cells and tissue \cite{trepat2007universal}.
\newline
\newline
\begin{acknowledgments}
AJL gratefully acknowledges support for this work from NSF Grant No. CMMI-1300514. MLG acknowledges support from ARO MURI (W911NF-14-1-0403), NSF Grant MCB-1344203 and U. Chicago MRSEC (DMR-1420709). SM thanks U. Chicago MRSEC for support through a Kadanoff-Rice fellowship. We thank William S Klug, Eric Dufresne and Daniel Blair for very helpful discussions.
\end{acknowledgments}

\section{References}
%
\section*{Methods}
\subsection*{Protein preparation}
Monomeric actin (G-actin) was purified using protocol adapted from Ref.\cite{spudich1971biochemical} from rabbit skeletal muscle acetone powder (Pel Freeze Biologicals, Product code: 41008-3). Small aliquots of actin solution were drop frozen in liquid nitrogen, and stored at -80°C. For each experiment fresh aliquots of frozen actin were thawed and used. FLN was purified from chicken gizzard and aliquots were also stored at -80°C.
\subsection*{In-vitro network formation}
10X actin polymerization-buffer (2 mM TrisHCl, 2 mM MgCl$_2$, 100 mM KCl, 0.2 mM DTT, 0.2 mM CaCl$_2$, 0.5 mM ATP, pH 7.5) was mixed with freshly thawed FLN.  Freshly thawed G-actin was mixed with 1X Ca-G-buffer (2mM Tris-HCl, 0.2mM ATP, 0.5mM DDT, 1mM NaN$_2$ and 0.1mM CaCl$_2$, pH 8). 1X Ca-G-buffer containing G-actin and 10X actin polymerization buffer containing FLN were mixed gently for 10 s just before loading the sample to initiate the polymerization process of actin filaments in the rheometer sample chamber.

\subsection*{Bulk rheology}
All rheological measurements were performed on a Bohlin Gemini HR Nano (Malvern Instruments) rheometer at a temperature of $25^0$C, using a 40 mm diameter acrylic plate and 160 $\mu$m gap. The acrylic plate has much lower moment of inertia compared to a similar metal plate and hence rheological measurements are more robust under fast direction switching of the plate rotation. We confirmed results were independent of plate geometry (Fig.S9). We also used a humidity chamber sealed with vacuum grease to reduce water evaporation from the sample during the rheology experiments and observed that the sample's rheological properties were essentially unchanged for periods up to $\sim$ 4 hours after loading the sample in the rheometer. The mechanomemory effect described in this manuscript is a robust effect, observed 100\% of the time in $>$ 35 independent samples.

After loading the sample, the build-up of linear elastic $G'$ and viscous $G''$ moduli was measured as a function of time at a fixed frequency $f$ = 0.5 Hz with a very small strain amplitude $\gamma_0$ = 0.02 (ensuring linear response) to monitor the polymerization process indicated by the increase in magnitude of $G'$ and $G''$ with time. After approximately 1 hour, the moduli saturate indicating a fully polymerized starting state of cross-linked actin network. This state is  the untrained state of the network. The tangent shear modulus $K (\gamma)$ is measured from the Lissajous plots (Fig.1d) by numerically calculating $\frac{\delta \sigma}{\delta \gamma}|_{\gamma}$ along the increasing strain magnitude under various training conditions. Memory is quantified by the asymmetry of tangent shear modulus $\Delta K$ in positive and negative direction at strain $|\gamma| = 0.4$, i.e. $\Delta K (|\gamma|) = K (+\gamma)$ - $ K (-\gamma)$. The choice of this strain controls the absolute value of the strength of memory, but does not affect changes in memory due to training or our analysis. The biggest source of error in shear modulus $K$ (or $\Delta K$) comes from the noise due to differentiation of high resolution data (we get $K$ by differentiating the stress with respect to strain). Here, we first numerically fit a spline smoothing curve (averaged over 10 adjacent points). Then we take the difference of this smooth curve from the raw data. The standard deviation (SD) of such differences gives the error in $K$. The error bars in $\Delta K$ are twice the standard deviations, since the uncertainty gets doubled when taking the difference.
\newline
\newline
To get the phase information, the complex tangent shear modulus $K^* (\sigma, \omega)$ was measured by superposing a small sinusoidal component of amplitude $\delta \sigma$ and angular frequency $\omega$ on a constant background stress $\sigma$ with $\delta \sigma = 0.1 \sigma$. Then we measured the amplitude of the sinusoidal strain $\delta \gamma$ at the same angular frequency $\omega$ and the phase difference $\Delta$ between the sinusoidal stress and the strain. The magnitude of the tangent shear modulus is given by 
$K (\sigma, \omega) = \frac{\delta \sigma}{\delta \gamma}|_{\sigma}$. 

Under steady shear, $P = \sigma \dot{\gamma}$ gives the instantaneous injected power per unit volume. However, since the system is visco-elastic, some part of this injected energy is dissipated and some part is stored. When the training stress is switched off, the system can relax back partially using the stored energy. We can also get insight about the dissipation in the system by superposing a small amplitude sinusoidal component on top of the large D.C. stress component. By this method we can measure the phase difference between the applied sinusoidal stress and resulting sinusoidal strain. When we plot the sinusoidal strain vs sinusoidal stress over a complete cycle, we get a closed curve (Lissajous plot) that can be well approximated by an ellipse in our case. The area under the curve is proportional to the total dissipated energy over a complete cycle. Here, it should be clear that the dissipated energy will also depend on the amplitude and frequency of the applied sinusoidal signals. Thus, by this method we cannot estimate the absolute power dissipation. However, if we use the same amplitude and frequency of the input sinusoidal component, we can compare how the power dissipation occurs as a function of time during the creep behavior under D.C. stresses. The energy dissipated over a complete cycle is $E_{cycl} = \pi\,\delta\sigma\,\delta\gamma\,sin\Delta$, leading to an average instantaneous power dissipation of $P(t) = (\omega /2\pi) E_{cycl}$.
 We define $P_{\alpha \beta}(t)$ to be the average instantaneous power dissipation in the $\beta$ direction in a sample trained in the $\alpha$ direction.
 Here $\alpha$ and $\beta$ can assume only the values $\pm$. Under steady shear, we calculate the injected power and subtract out the recoverable power at every instant to estimate the time dependent power dissipation (Fig.S2). However, to do this one has to assume that the injected power at the start up of creep response corresponds to the recoverable power at the start of recovery curve.
\subsection*{Confocal microscopy}
20\% molar ratio of labeled actin monomers (actin from rabbit skeletal muscle conjugated with Alexa fluorophore 568, Life Technologies Corp., product part no. A12374) was added to 80\% G-actin and then the combination is mixed gently with 1X Ca-G-buffer. Ca-G-buffer con-taining both labeled and unlabeled G-actin and 10X actin polymerization buffer containing FLN were mixed gently for 10 s just before loading the sample into a home built glass chamber (thickness $\sim 100 \mu$m) mounted on the sample stage of a laser scanning confocal micro-scope (Nikon) with a 60X water immersion objective. Z-stack images were recorded approximately after 1 hour of polymerization.
\section*{Model}
We developed a mathematical model of the system as a two-dimensional network of $N$ semiflexible filaments with identical elastic properties and 
length $L$. To construct the numerical network, each filament is placed at random in a box of area $A$ with an orientation picked from a distribution $P(\theta)$, where $\theta$ denotes the angle between the filament and the nematic director $\mathbf{n}$ of the network.
\subsection*{Single filament and crosslink mechanics}
Using classical Bernoulli-Euler beam theory, the mechanical properties of the filaments are characterized by the linear stretching and bending moduli, 
$\mu$ and $\kappa$ respectively. The elastic length scale or bending length $\lambda_b =\sqrt{\kappa/\mu}$ determines the relative contribution of the filament's bending and stretching modes to the total energy. The thermal persistence length of the filaments is given by $\ell = \kappa/k_{{\rm B}}T$. The energy of a single filament can then be written as
\begin{eqnarray}
\label{eq:strain_energy}
E_{\mathrm{f}}&=&\int_{L}ds\left[\frac{\mu}{2}\left|\frac{d\mathbf{r}}{ds} \right|^{2}+\frac{\kappa}{2}\left(\frac{d\theta}{ds} \right)^{2} \right],
\end{eqnarray}
where we used an arclength parametrization of the filament with $\mathbf{r}(s)$ the position of the filament cross-section after deformation, and $\theta$ the rotation of the filament caused by that deformation.
\newline
\newline

We implement a mechanical model for the filamin cross links that is based on previous experimental measurements \cite{xu2013domain}.  
The filamin contour length is $\sim$100 nm, but is comprised of subunits of rest length $\sim$10 nm that unfold under force.  Previous experimental results 
show that, during unfolding, the subunits follow a worm-like chain model \cite{xu2013domain}. In the simulation, the saw-tooth force-extension curve is reduced to two regimes, reflecting the regime of subsequent unfolding and the nonlinear stiffening observed during stretching of the last segment.  This was adopted in order to reduce problems arising from 
numerical instability. For lengths below 150 nm, the 1d spring constant is $\mu_{lc}=100pN$. In the stiffening regime above $150nm$, there is a linear increase in the 
spring constant of 10 pN/nm for further extension.
 
The geometry of the network is characterized by the scalar amplitude of the nematic order parameter $S$, found from the width of the 
distribution $P(\theta)$:
\begin{eqnarray}
S &=& 2\int_{0}^{\pi}d\theta P(\theta)\cos(2\theta).
\end{eqnarray}
Our results are remarkably insensitive to the exact form of distribution of filament angles (Fig.S10).

We have been using the following distribution in our simulations:
\begin{eqnarray}
P(\theta) &=& \frac{e^{{\alpha \cos{2\theta}}}}{2\pi I_{0}(\alpha)}
\end{eqnarray}
where $I_0(\alpha)$ the modified Bessel function of the first king of order 0. The parameter $\alpha$ is tied to the nematic angle and is found numerically by solving $S=I_{1}(\alpha)/I_{0}(\alpha)$.
\newline
\newline
Another geometric characteristic of the network is the mean distance between consecutive crosslinks along a given filament, $l_c$, which is inversely proportional to the network density and depends on the nematic order parameter $S$. This quantity is necessary to assess the propensity for buckling. Long spans between 
consecutive cross links buckle at a lower compressive stress than shorter ones. This quantity may be computed by considering a filament lying at an 
angle $\theta$ with respect to the nematic director. The probability of a second filament crossing at an angle $\psi$ with respect to the first is given 
by $L^2 \left|\sin\psi\right|/A$. In order to obtain the crossing probability independent of the relative angle $\psi$, we need to integrate the above quantity over $\psi$ with a weight $P (\theta+ \psi)$. Doing this gives:
\begin{eqnarray}
P_{\mathrm{cross}}(\theta) &=& \frac{2L^2}{A}\int_{0}^{\pi}d\psi \sin{\psi} P (\theta+ \psi)
\end{eqnarray}
This quantity is simply the probability that a given filament oriented with angle $\theta$ to the nematic direction is crossed by a second filament with its direction chosen from the angular probability distribution and its position chosen randomly. We will again assume that our system contains a large number of filaments $N$ in a large area $A$, so that $P_{\mathrm{cross}}(\theta)$ is small; in this limit, the probability distribution $p_n(\theta)$ for the number of crosslinks on a filament oriented with angle $\theta$ to the nematic direction can be approximated by the following Poisson distribution:
\begin{eqnarray}
p_{n}(\theta)  &=& \frac{e^{-n_{c}(\theta)}[n_{c}(\theta)]^{n}}{n!}
\end{eqnarray} 
where $n_c(\theta)$ is the mean number of cross links on a filament oriented with angle $\theta$, and is written:
\begin{eqnarray}
n_{c}(\theta) &= & 2\rho L^2\int_{0}^{\pi}d\psi \sin(\psi)P(\theta+\psi),
\end{eqnarray}
with $\rho = N/A$ the filament density. We then calculate the distance $\Lambda_c$ between the two most distance crosslinks on a filament  and the average number of intervals between two crosslinks on a filament $N_c$. These quantities are given by:
\begin{eqnarray}
\Lambda_c &=& 2NL\int_{0}^{\pi}d\theta P(\theta)\sum_{n=0}^{\infty}p_n(\theta)\frac{n-1}{n+1}\\
N_c &=& 2N\int_{0}^{\pi}d\theta P(\theta)\sum_{n=2}^{\infty}p_n(\theta)(n-1)
\end{eqnarray}
Finally, the ratio of $\Lambda_c$ and $N_c $ yields the mean distance between crosslinks: 
\begin{equation}
\label{eq:lc}
l_{c} = L\frac{1+\left\langle e^{-n_c} \right\rangle -2\left(\left\langle
\frac{1}{n_c}\right\rangle  -\left\langle \frac{e^{-n_c}}{n_c} \right\rangle
\right)}{\left\langle n_c \right\rangle - \left( 1- \left\langle e^{-n_c}
\right\rangle \right)},
\end{equation}
where $\left\langle \cdot \right\rangle$ denotes the averaging operation
over $\theta$ with weight $P(\theta)$. 

\begin{figure}
	\centering
	\includegraphics [width = 8 cm]{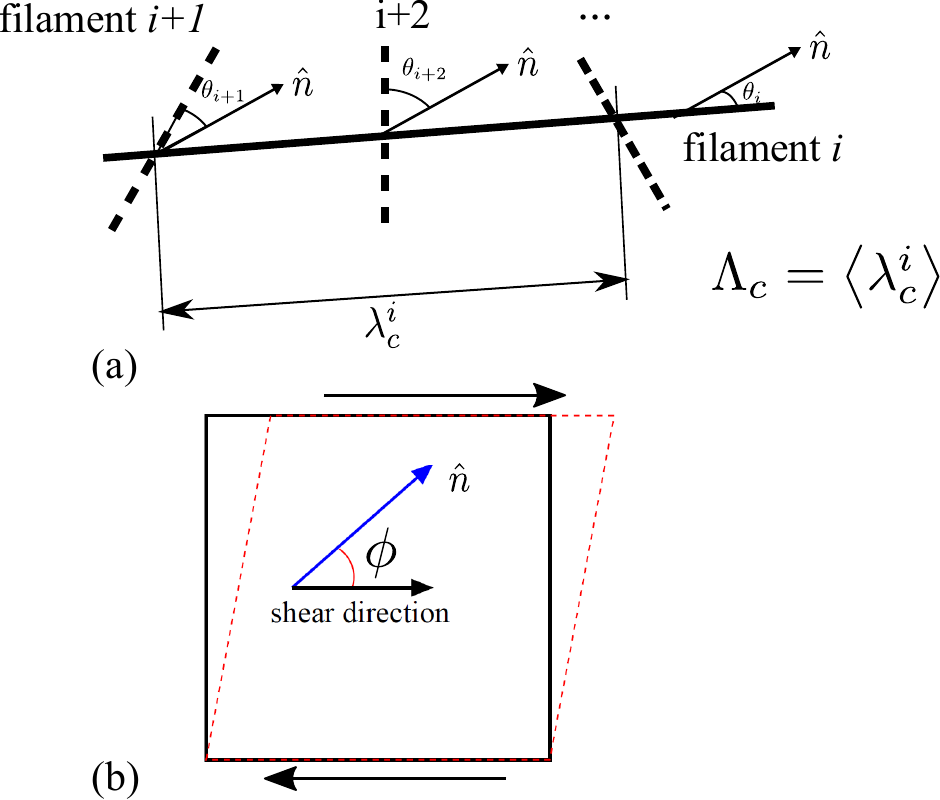}
	\renewcommand{\thefigure}{M1}
	\caption{\textbf{Geometric properties of the network.} (a) $\Lambda_c$ is the average (over filaments) of the distance $\lambda_{c}^{i}$  between the two 
most distance crosslinks on a filament $i$. The orientation of filament $i$ is defined as the angle $\theta_{i}$ made with the nematic director $\hat{n}$. 
(b) The nematic director $\hat{n}$, in turn, makes an angle $\phi$ 
with respect to the displacement direction of the applied shear (here the $x$ axis).}
	\label{M1}
\end{figure}

\subsection*{Simulation details}
\noindent
We consider a system  composed of $N$ semiflexible filaments connected together by cross links at points of intersection. 
The cross links are extensible nonlinear elastic elements as described above, but they supply no constraint torques, being free to rotate. The filaments 
are spatially discretized by placing nodes at each cross link, as well as at regular intervals between adjacent cross links, which ensures a 
sufficient level of spatial refinement for numerical convergence. Typically, networks are generated with $4000$ filaments 
discretized into $100~000$ finite elements. The construction of each network is based on the two geometric parameters defined 
above: (a) the nematic order parameter $S$ and (b) the mean filament density $l_c$. To build the network, we add filaments sequentially to the simulation box 
with typical dimensions of $A = 64L^2$.  We numerically evaluate the strain energy using the discretized version of equation (\ref{eq:strain_energy}):
\begin{eqnarray}
\tilde{\mathcal{H}} &=& \frac{\mu}{2}\sum_{\mathrm{segments}}\frac{(d - d_0)^2}
{d_0} + \kappa \sum_{\mathrm{angles}}\frac{1-\cos{\beta}}{l_0},
\end{eqnarray}
where $d$ and $d_0$ are respectively the current and original length of the
segments, and $l_0$ is the average rest length of two adjacent segments
forming an angle $\beta$. The nematic order parameter is set by choosing the direction of the filaments from a Gaussian distribution with a width determined to give 
the desired nematic order parameter $S$.  The centers of the mass of the filaments remain uniformly distributed during the construction of all networks. 
We have explored the mechanics of networks constructed using other anisotropic orientational distributions of the filaments, and found only 
small changes in the collective mechanics of the networks when both the network density and nematic order parameter were held constant. See Fig.S10.

We use the Lees-Edward method~\cite{Search1972} to simulate the shearing deformation of the network with
periodic boundary conditions, and the static equilibrium of the system is
obtained by energy relaxation using a quasi-Newton minimization algorithm~\cite{Byrd1994,Zhu1994}.

\subsection*{From residual strain to nematic order}
\noindent
The residual strain measured after training implies a realignment of the filaments in the network. Assuming that the residual strain results from the 
affine deformation of a formerly isotropic network, we compute the nematic order parameter $S$ induced by training. Choosing an arbitrary filament from an 
isotropic 2D network of filaments (length $L$), we write its initial orientation as 
\begin{eqnarray}
\mathbf{R} &=& L\left[\begin{array}{c}
\cos \theta\\
\sin \theta
\end{array}\right]
\end{eqnarray}
with $\theta$ the angle made by the filament with respect to the $x$ axis. We apply a simple shear 
along direction $x$ in the $xy$ plane. Under this deformation the filament rotates make the angle $\theta'$ with respect to the same $x$ axis, given by
\begin{eqnarray}
\mathbf{R}' &=& L\left[\begin{array}{c}
\cos \theta + \gamma\sin \theta\\
\sin \theta
\end{array}\right] \;=\; L\left[\begin{array}{c}
\cos \theta' \\
\sin \theta'
\end{array}\right].
\end{eqnarray}
The new angle $\theta'$  is thus 
\begin{eqnarray}
\label{mapping}
\theta' &=&\arctan \left[\frac{\sin \theta}{\cos \theta + \gamma \sin \theta}\right].
\end{eqnarray}
\noindent
Considering the ensemble of filament orientations given by the unit vectors $\mathbf{u} = (\cos \theta,\sin \theta)$ in the isotropic network, we apply the mapping 
Eq.~\ref{mapping} to obtain the new (anisotropic) distribution $\mathbf{u}' = (\cos \theta',\sin \theta')$ of filament orientations after the shear deformation  $\boldsymbol \lambda =\boldsymbol \delta + \gamma \mathbf{x}\mathbf{y}$. From this distribution we directly compute the resulting nematic order parameter tensor by averaging over the initially isotropic distirbution of angles $\theta$.  Denoting this average by $\left\langle .\right\rangle $,  we write the order parameter $S_{ij}=\left\langle u_{i}' u_{j}' \right\rangle-\frac{1}{2}\delta_{ij}$ as:
\begin{eqnarray}
S_{11} &=& \frac{\gamma^2 + 2}{\gamma^2 + 4}-\frac{1}{2},\;\;S_{12}=S_{21}\;=\;\frac{\gamma}{\gamma^2+4}\nonumber\\
S_{22}&=&\frac{2}{\gamma^2+4}-\frac{1}{2},
\end{eqnarray}
Finally, after diagonalization, we find the magnitude of the order parameter $S$ to be given by 
\begin{equation}
S(\gamma) =\frac{\gamma}{\sqrt{\gamma^2+4}}
\end{equation}
This provides an explicit expression for the nematic order as a function of the residual strain $\gamma$.
\subsection*{Mean field approach}
\noindent
We develop a mean field calculation of the shear modulus as a function of nematic order. We shear the network along the $+x$ direction, where the 
nematic director makes an angle $\phi$ with respect to $x$ (Fig.M1b). We assume that  filaments buckle only when under a compressive load larger than the Euler buckling 
load $p_c$, which depends on the length of the filament segment between consecutive cross links and on the angle that that  filament 
makes with the shearing direction. We further assume that one may neglect the (small) amount of elastic energy stored in filaments post-buckling. 

\noindent
Under these assumptions, a filament oriented making an angle $\psi$ with the $x$-axis has an energy per unit length
\begin{equation}
\label{energy-per-length}
E_{\mathrm{f}}(\psi)=\mu \gamma^2 \cos^2\psi \sin^2\psi/2,
\end{equation}
if not buckled. Based on the Euler buckling criterion, there is a wedge of width 
$2 \omega_{\rm Euler}(\ell)$ around the compression direction of the shear  so that filament segments of length greater than $\ell$ oriented within this wedge will buckle. 
We wish to exclude such buckled filaments from the computation of the elastic energy stored in the network using Eq.~\ref{energy-per-length}.  We replace this 
segment-length dependent quantity $2 \omega_{\rm Euler}(\ell)$ with an averaged excluded wedge angle, $\theta_{c}$.   This is determined self-consistently 
by setting the average filament segment length (averaged only over the wedge of half angle $\theta_{c}$ to be equal to the load-dependent critical length for 
Euler buckling $l_{\mathrm{buckle}}(\gamma)$. This results in a self-consistent integral equation to be solved for $\theta_{c}$, and which depends on both the degree of
nematic order and shear strain $\gamma$:
\begin{equation}
\label{eq:lc}
l(\theta_c) = L\frac{1+\left\langle e^{-n_c} \right\rangle -2\left(\left\langle
\frac{1}{n_c}\right\rangle  -\left\langle \frac{e^{-n_c}}{n_c} \right\rangle
\right)}{\left\langle n_c \right\rangle - \left( 1- \left\langle e^{-n_c}
\right\rangle \right)}=l_{\mathrm{buckle}},
\end{equation}
where $\left\langle \cdot \right\rangle$ notes denotes the average over the wedge for a given degree of nematic order $P(\theta)$.  This is solved numerically. 

\noindent
We then compute the total network energy by integrating Eq.~\ref{energy-per-length} over the filament orientation and length~\cite{Missel2010}, excluding 
wedge of half angle $\theta_{c}$ surrounding the compression direction, oriented along $\theta_{\rm comp}$= -45 degrees w.r.t. the shearing direction. The total energy of the network is
\begin{eqnarray}
E(\phi) &=& \mu\rho\gamma^2\left(\int_{0}^{\theta_{\mathrm{comp}}-\theta_{\mathrm{c}}} \cos^2\psi \sin^2\psi/2 \right.\nonumber \\
&\times&\left. \sum_{n=2}^{\infty}\frac{L(n-1)p_{n}(\psi-\phi)}{n+1}\right.\nonumber\\
&+&\left. \int_{\theta_{\mathrm{comp}}+\theta_{\mathrm{c}}}^{\pi}\cos^2\psi \sin^2\psi/2\right. \nonumber\\ 
&\times & \left.\sum_{n=2}^{\infty}\frac{L(n-1)p_{n}(\psi-\phi)}{n+1}\right),
\end{eqnarray}
Using the fact that $E(\phi)=K(\phi)\gamma^2/2$ for small shear, we write the tangent modulus as
\begin{eqnarray}
\label{MFT-tangent-modulus}
K(\phi) &=& 2\mu\rho \int_{0}^{\theta_{\mathrm{comp}}-\theta_{\mathrm{c}}}d\psi \left(\cos^2\psi \sin^2\psi/2 \right. \nonumber \\
& \times & \left. \sum_{n=2}^{\infty}\frac{L(n-1)p_{n}(\psi-\phi)}{n+1}\right)\nonumber\\
&+& 2\mu\rho \left. \int_{\theta_{\mathrm{comp}}+\theta_{\mathrm{c}}}^{\pi}d\psi\left(\cos^2\psi \sin^2\psi/2\right.\right. \nonumber \\
& \times & \left. \sum_{n=2}^{\infty}\frac{L(n-1)p_{n}(\psi-\phi)}{n+1}\right).
\end{eqnarray}
The mean-field value of $\Delta K$ computed using Eq.~\ref{MFT-tangent-modulus} is shown in Fig.~\ref{F4}d as a dashed black line.  
The mean-field model is consistent with both the simulation (red line and symbols) and the experiment (points).\\
\newpage
\begin{table*}[b]
\centering
\caption{Table of parameters} 
 \begin{tabular}{c c c c} 
 \hline
 Parameters & Symbol & magnitude & Ref. \\ [0.5ex] 
 \hline
 (Geometrical)\\
 F-actin length                   								& $L$ & 2 $\mu$m &Viamontes \textit{et al.}, PRE, 2006 \cite{Viamontes2006}\\ 
Average distance between cross-linking points								&  &  $ \approx80$ nm & \\ 
Persitence length of filamin subunits								&  &  $ \approx1$ nm & Xu \textit{et al.}, Biophysical J., 2013 \cite{xu2013domain}\\
 Nematic order                                  & $S$  & 0-1 & \\       
  Finament density                                             &  & 16 filaments / ($\mu m)^2$ &  \\
 [1ex] 
 \hline \hline 
(Mechanical)\\                                              
F-actin bending mod.                                        & $\kappa$ & 6.5$\times 10^{4}$ $pN.nm^2$ & Yanagida et al., Nature, 1984 \\ 
F-actin 1d modulus                                  & $\mu$ & 17.1 nN & Kojima et al., PNAS, 1994 \cite{Kojima1994}\\ 
F-actin Euler buckling force (L = 2 $\mu$m)                 & & 0.16 $pN$   & Calculated using `L' and `$\kappa$' values\\
[1ex] 
\hline
 \end{tabular}
\end{table*}

\begin{figure*} [b]
	\centering
	\includegraphics [width= 9 cm]{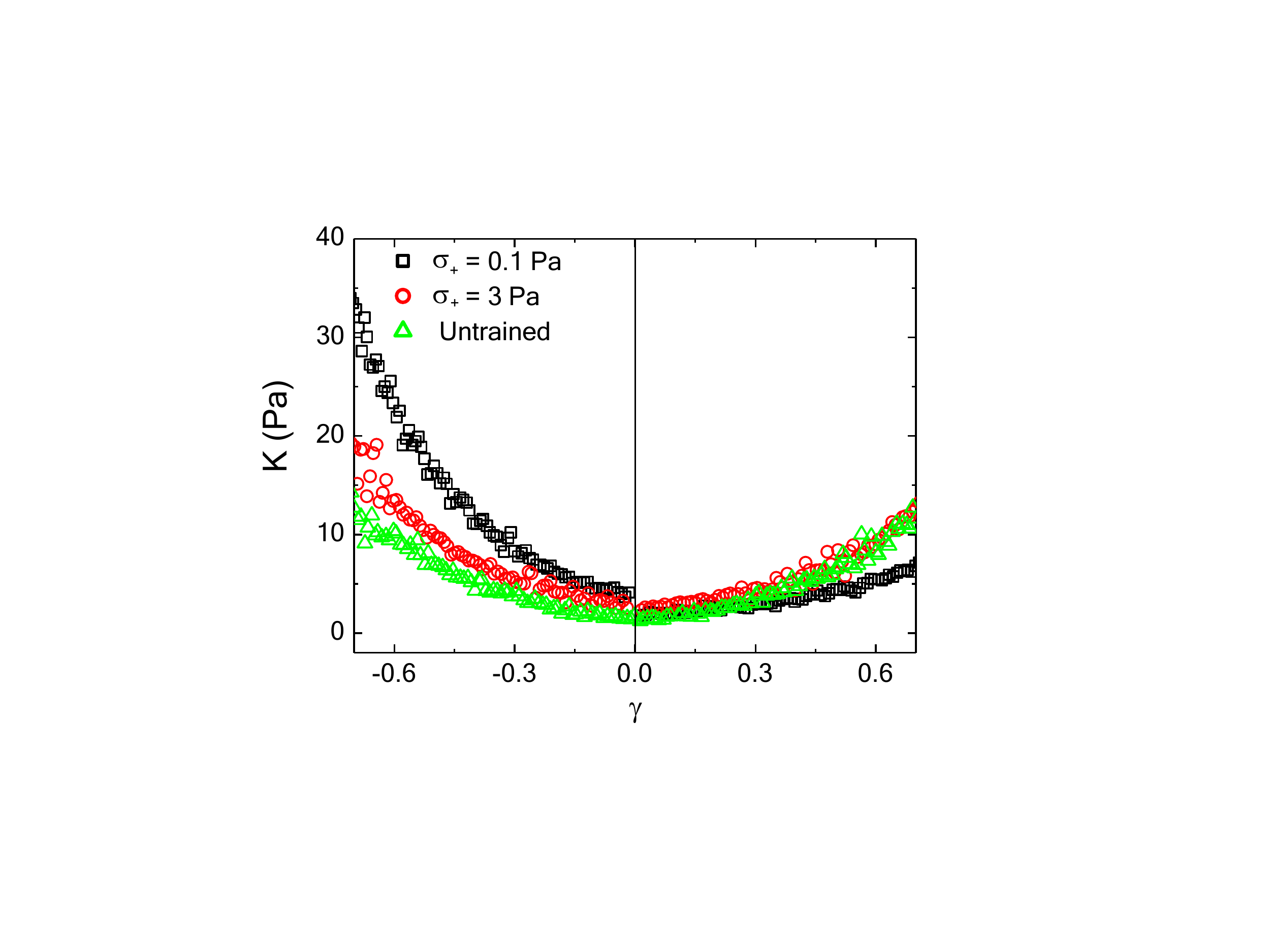} 
	\renewcommand{\thefigure}{S1}
	\caption{\textbf{Bringing a maximally trained sample back to the untrained state.} Raw data for the shear modulus $K$ as a function of applied strain $\gamma$ for two training stress values (indicated in the figure legend) applied in the opposite direction to the initial training as mentioned in Fig.2d in the main text. This plot indicates that, starting with a highly asymmetric state (large memory), it is possible to bring back the sample to a state that is close to the untrained (no memory) state by applying suitable opposing stress pulses.  
 }
	\label{FS1}
\end{figure*}	
\begin{figure*} [b]
	\centering
	\includegraphics[width=16 cm]{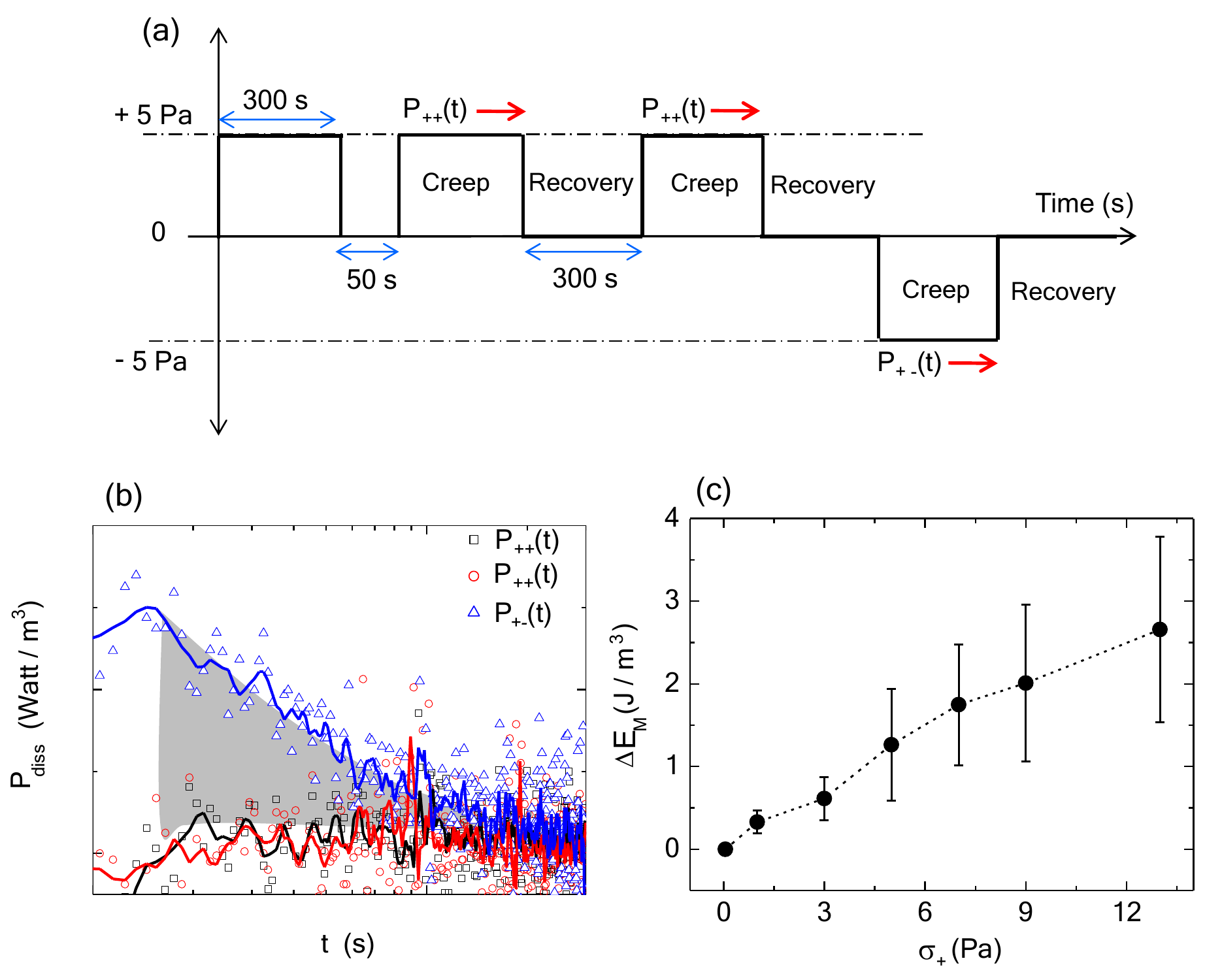}
	\renewcommand{\thefigure}{S2}
	\caption{\textbf{Absolute energy dissipation in re-writing a memory from creep and recovery measurements.} \textbf{a}, Schematic of the protocol: An untrained sample ($c_a$ = 24 $\mu$M actin + 5 \% FLN) is first trained by $\sigma_+$ = 5 Pa stress in the positive direction for 300 s. After waiting for 300 s, a stress of magnitude 5 Pa is again is applied for 300 s and then the stress is switched off ($\sigma$ = 0 Pa) for 300 s. We estimate the shear rate as a function of time by numerically differentiating the strain data during both creep ($\dot{\gamma}_c(t)$) and recovery ($\dot{\gamma}_r(t)$) as indicated in Fig.3 in the main text. The difference shear rate is measured by $\Delta\gamma (t) = \dot{\gamma}_c(t) - \dot{\gamma}_r(t)$. The quantity $\sigma\,\Delta\dot{\gamma}(t) $, gives the instantaneous power dissipation. $P_{++}(t)$ and $P_{+-}(t)$ represent the power dissipation when the creep stress direction is same or opposite to the training direction, respectively.  \textbf{b}, $P_{++}(t)$ and $P_{+-}(t)$ as a function of time. The integrated area between $P_{+-}(t)$ and $P_{++}(t)$ indicated by the gray region gives the total dissipated energy $\Delta\,E_{M}$ in re-writing a memory. The symbols represent the raw data and the lines represent the spline curve (obtained by 5 points moving averaging). \textbf{c}, $\Delta\,E_{M}$ as a function of D.C. stress pulse magnitude. The error bars represent the standard deviation of the fluctuations in power vs the time data in (\textbf{c}).}
	\label{FS2}
\end{figure*}

\begin{figure*} [h]
\centering
\includegraphics[width= 16 cm]{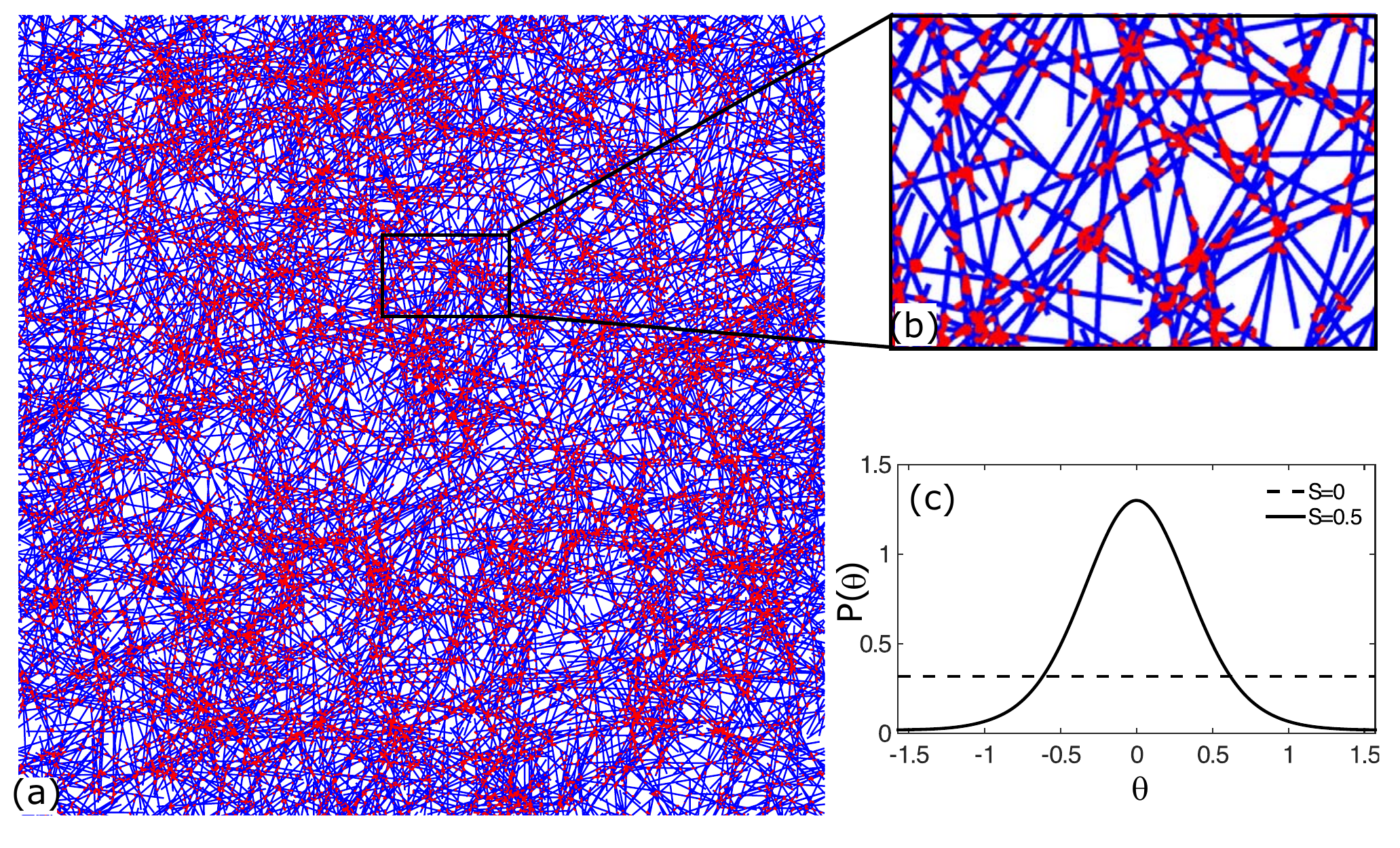}
\renewcommand{\thefigure}{S3}
\caption{\textbf{Example of a semiflexible network in two dimensions.} \textbf{a}, Actin filaments are in blue and the filamin cross-linkers in red. \textbf{b}, The close-up view of the cross-linking of filaments shown in (\textbf{a}). \textbf{c}, The Gaussian orientation distribution of filaments in the nematic network. Here, the orientation angle $\theta$ of a filament is measured w.r.t. the nematic director of the network.}
\label{FS3}
\end{figure*}
\begin{figure*} [h]
	\centering
	\includegraphics [width= 10 cm]{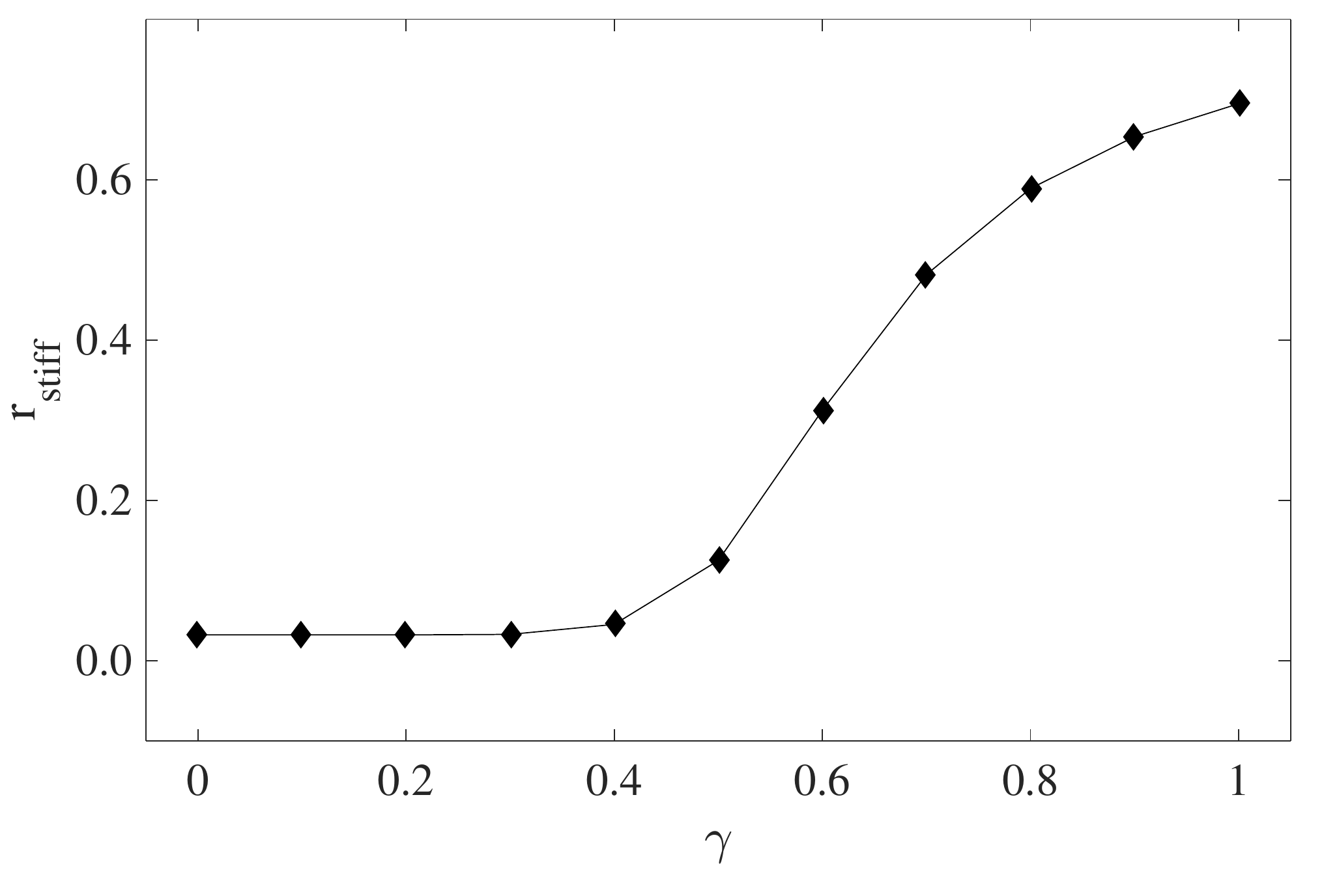} 
	\renewcommand{\thefigure}{S4}
	\caption{\textbf{Stiffening of filamin cross-linkers obtained by numerical simulation.} Fraction of filamin molecules getting stiffened when the cross-linked actin network (e.g. Fig.S3) is strained. We see that, beyond a strain of $\sim$ 0.4, the number of stiffened filamin molecules increases rapidly indicating that constitutive non-linearity takes over the geometric non-linearity at large strain values.
 }
	\label{FS4}
\end{figure*}

\begin{figure*} [h]
	\centering
	\includegraphics [width= 16 cm]{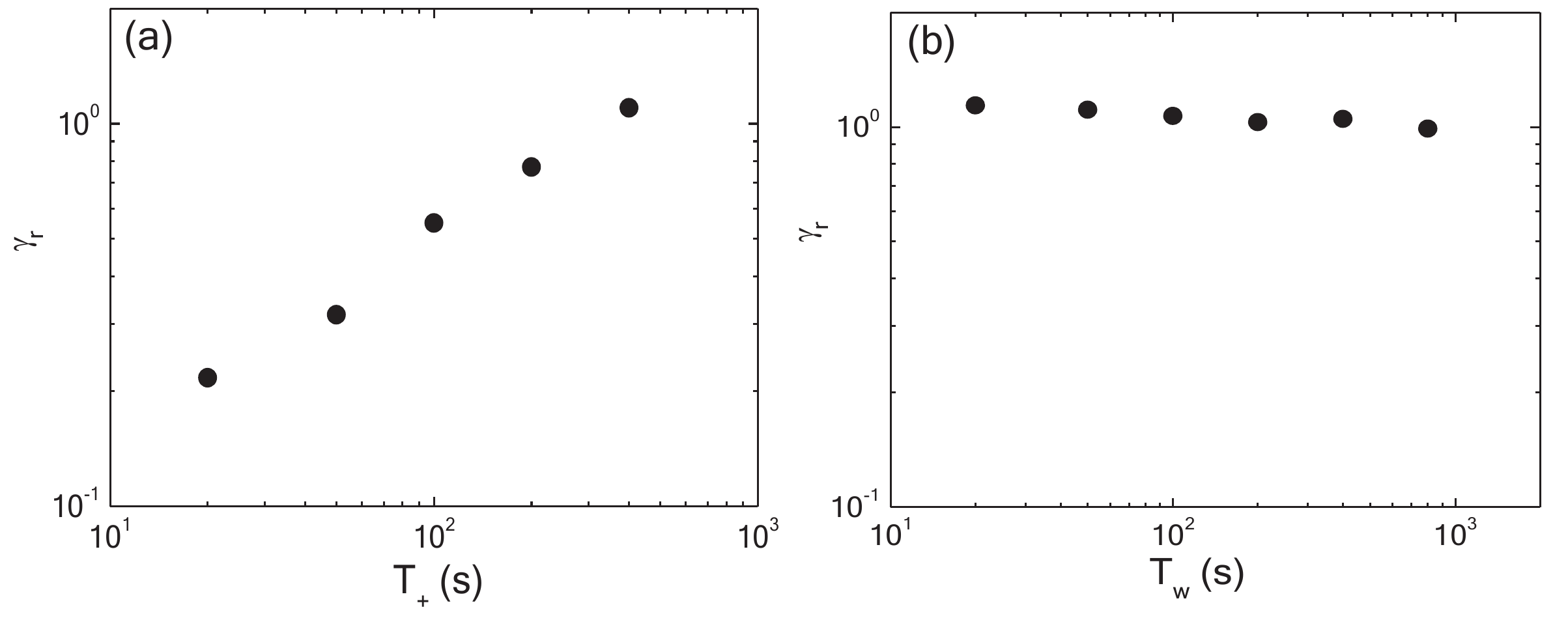} 
	\renewcommand{\thefigure}{S5}
	\caption{\textbf{Evolution of residual strain as a function of training and waiting time.} \textbf{a}, An untrained sample ($c_a$ = 24 $\mu$M actin + 5 \% FLN) is trained by $\sigma_+$ = 4 Pa stress in the positive direction with increasing training time $T_+$.
The cumulative residual stain is shown as a function of $T_+$. \textbf{b}, The residual strain as a function of waiting time. The sample is initially trained with $\sigma_+$ = 4 Pa applied for $T_+$ = 300 s.}
	\label{FS5}
\end{figure*}

\begin{figure*} [h]
	\centering
	\includegraphics [width= 13 cm]{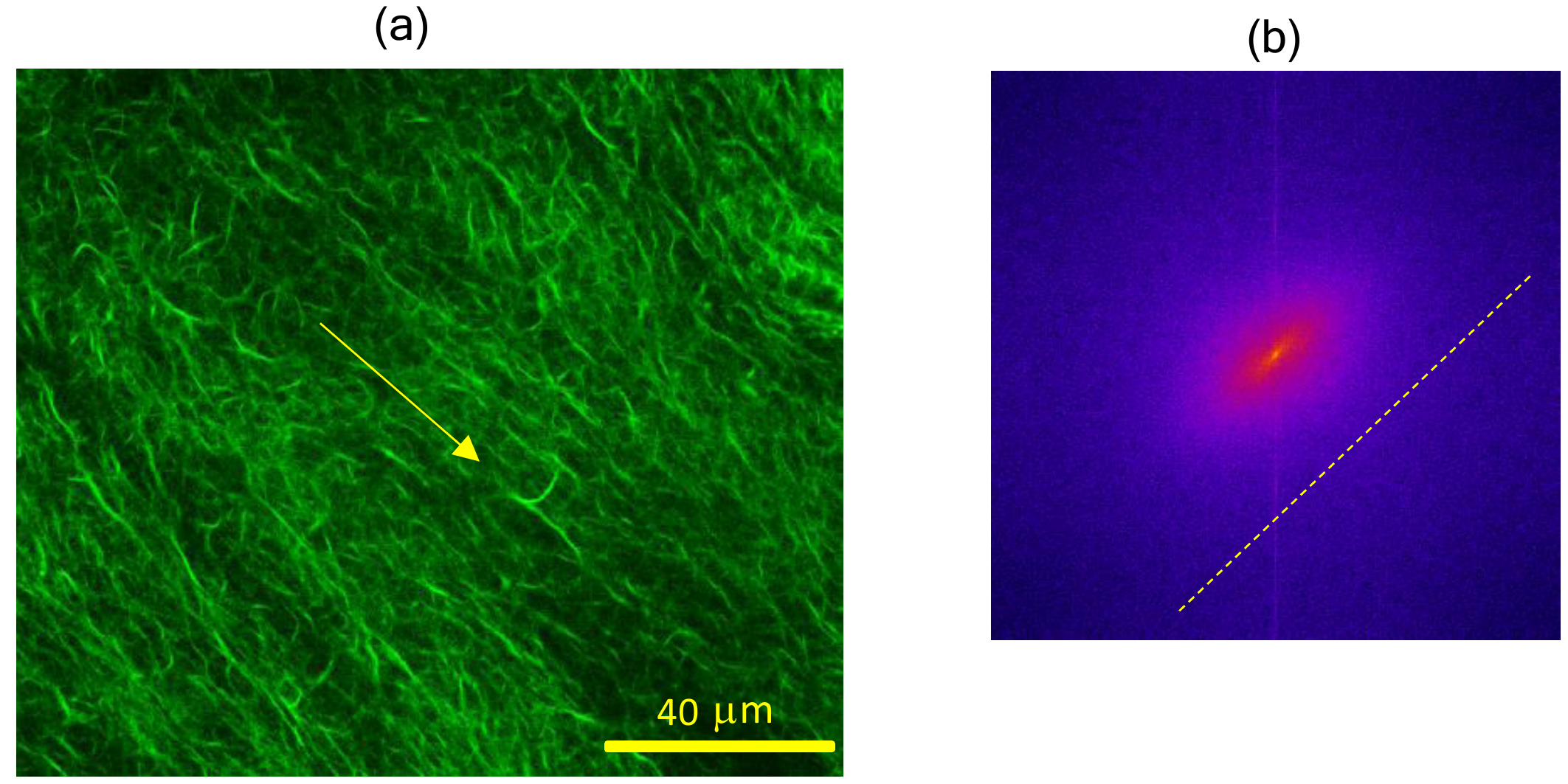}
	\renewcommand{\thefigure}{S6}
	\caption{\textbf{Network structure under shear deformation: frozen nematic order.} A freshly polymerized sample ($c_a$ = 12 $\mu$M actin + 5 \% FLN) is first deformed to a maximum strain of 1 at a rate 0.1 $s^{-1}$. After a waiting time of 500 s at the maximum strain, 20 confocal image planes are recorded over a vertical height of 20 $\mu$m. \textbf{a}, shows one such slice and \textbf{b}, represents the corresponding FFT image where the colors from blue to red to bright yellow represent increasing magnitude of Fourier amplitudes. We see a frozen in nematic order in the direction of applied shear (shown by an arrow), resulting from the residual strain.
 }
	\label{FS6}
\end{figure*}

\begin{figure*} [h]
	\centering
	\includegraphics [width= 10 cm]{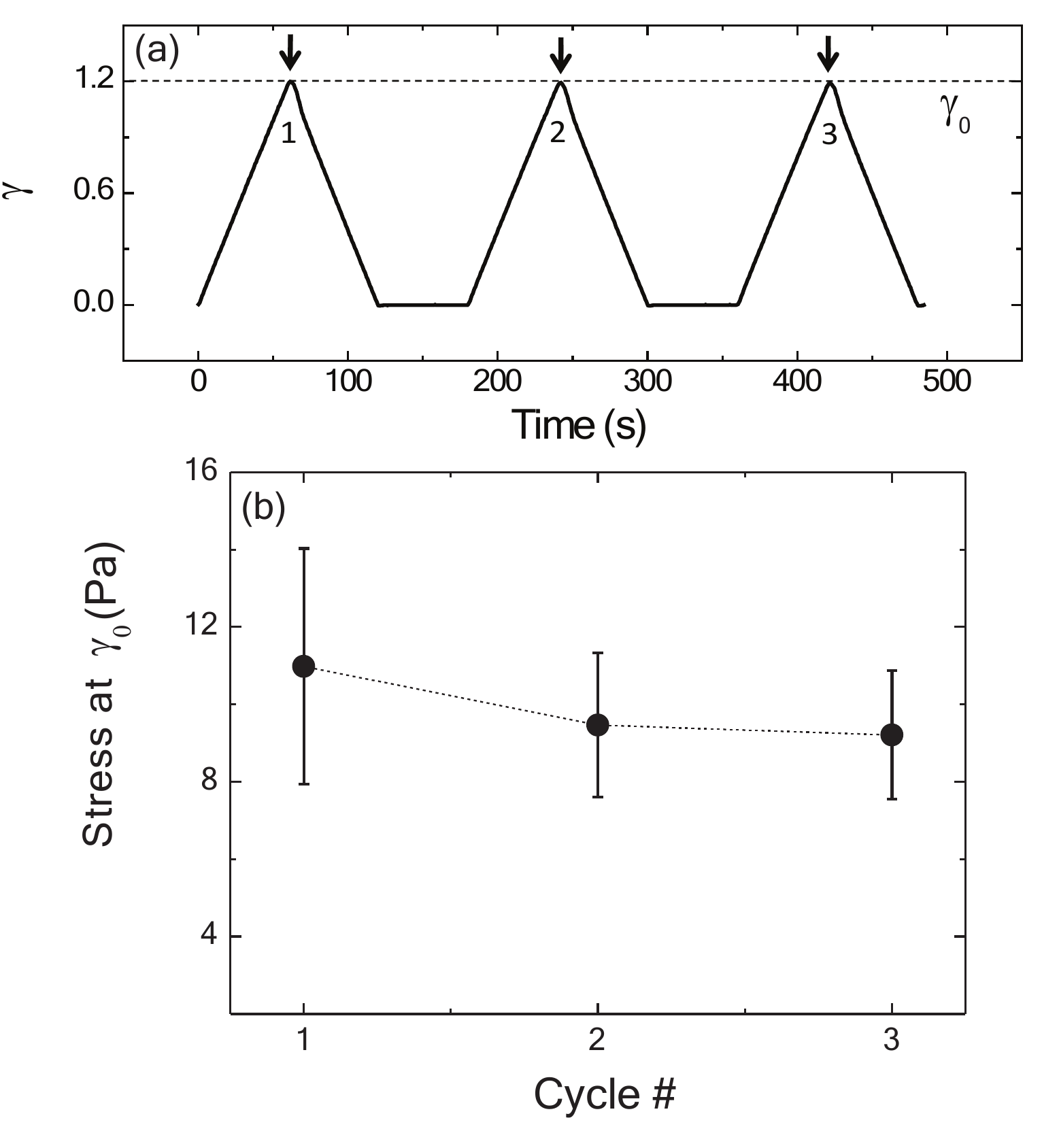} 
	\renewcommand{\thefigure}{S7}
	\caption{\textbf{Variation of peak stress under cyclic loading.} \textbf{a}, Schematic of the cyclic loading protocol. The strain ramp rate is 0.02 $s^{-1}$ and the waiting time between two consecutive pulses is 60 s. The maximum strain position $\gamma_0$ is indicated by the arrows. \textbf{b}, Stress values at the position of maximum strain for three consecutive strain cycles as indicated in \textbf{a}. The error bars are estimated from the standard deviation of the stress values from three independent experiments on the same sample ($c_a$ = 24 $\mu$M actin + 5 \% FLN).  
 }
	\label{FS7}
\end{figure*}

\begin{figure*} [h]
	\centering
	\includegraphics [width= 17 cm]{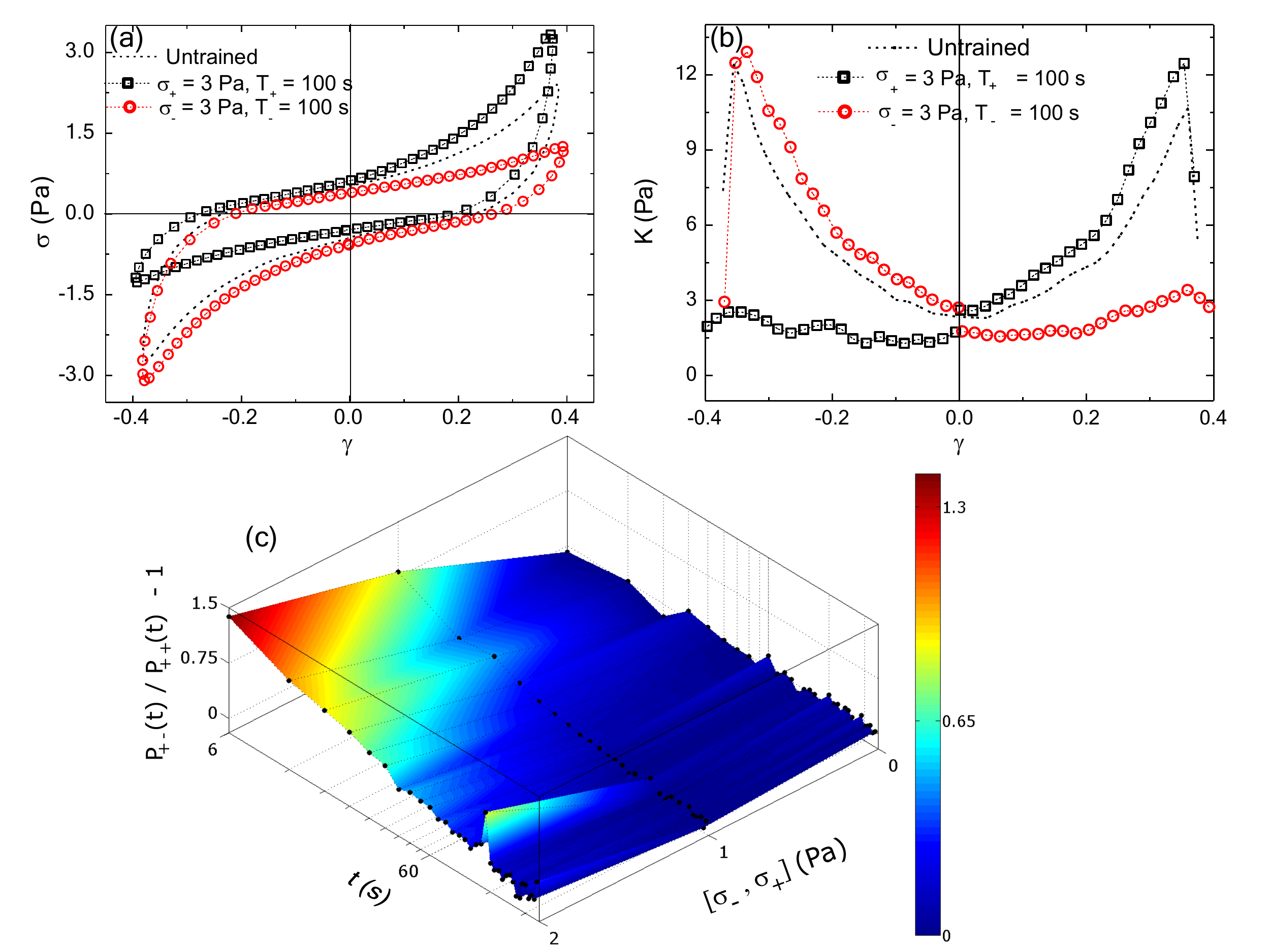} 
	\renewcommand{\thefigure}{S8}
	\caption{\textbf{Memory effect in actin network cross-linked with a rigid cross-linker $\alpha$-actinin.} \textbf{a}, Lissajous plots (stress vs. strain) for an untrained, positively trained and negatively trained sample ($c_a$ = 24 $\mu$M actin + 5 \% $\alpha$-actinin). 
The training stress magnitude and duration in different cases are shown in the figure legend. \textbf{b}, Shear modulus as a function of strain obtained from the Lissajous plots in panel \textbf{a}. \textbf{c}, Normalized time dependent power dissipation upon stress reversal, estimated by superposing a small A.C. component of stress on the D.C. stress pulses (see Fig.2d and Methods). The 3-D plot is generated by varying the magnitude [$\sigma_+$, $\sigma_{-}$] of D.C. pulses.}
	\label{FS8}
\end{figure*}

\begin{figure*} [h]
	\centering
	\includegraphics [width= 17 cm]{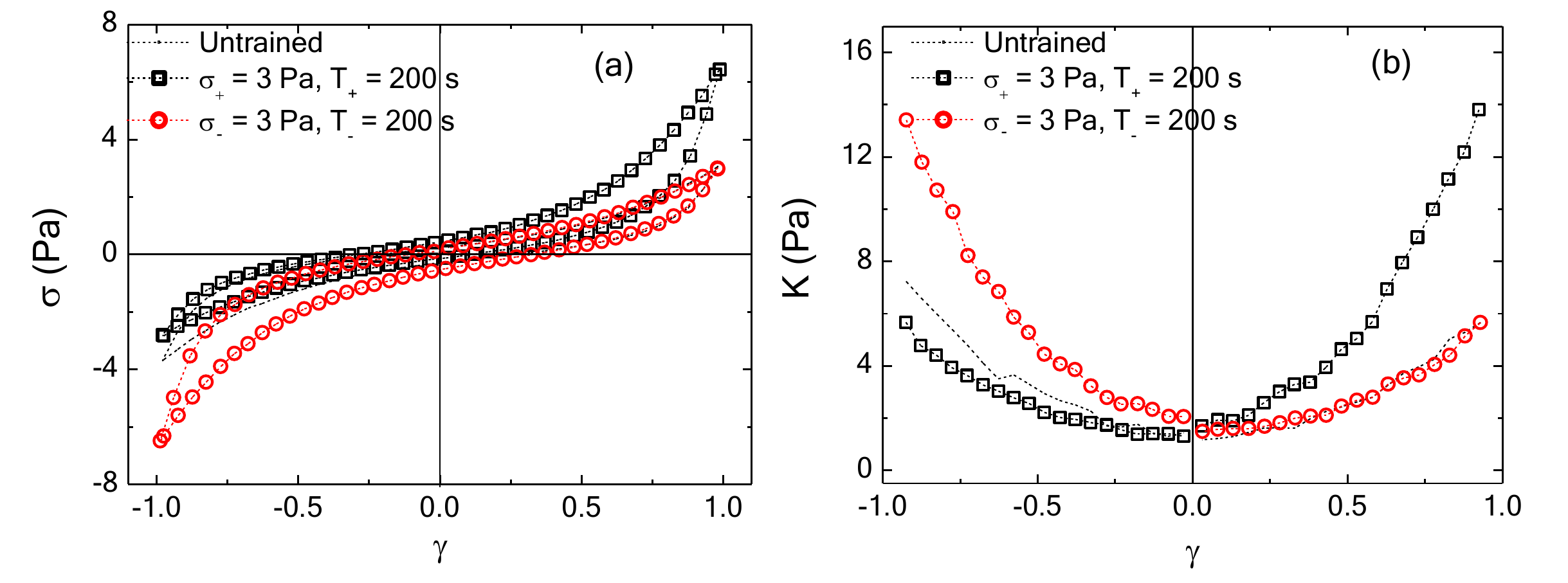} 
	\renewcommand{\thefigure}{S9}
	\caption{\textbf{Memory effect in actin network probed by a cone and plate rheometer.} \textbf{a}, Lissajous plots (stress vs. strain) for an untrained, positively trained and negatively trained sample ($c_a$ = 24 $\mu$M actin + 2.5 \% FLN). The training stress magnitudes and durations are indicated in the figure legend. \textbf{b}, Shear modulus as a function of strain obtained from the Lissajous plots in panel \textbf{a}. These results are similar to that obtained using a parallel plate rheometer. Here, the diameter of the cone is 20 mm and the cone angle is 1$^{\circ}$.
 }
	\label{FS9}
\end{figure*}

\begin{figure*} [h]
	\centering
	\includegraphics [width= 10 cm]{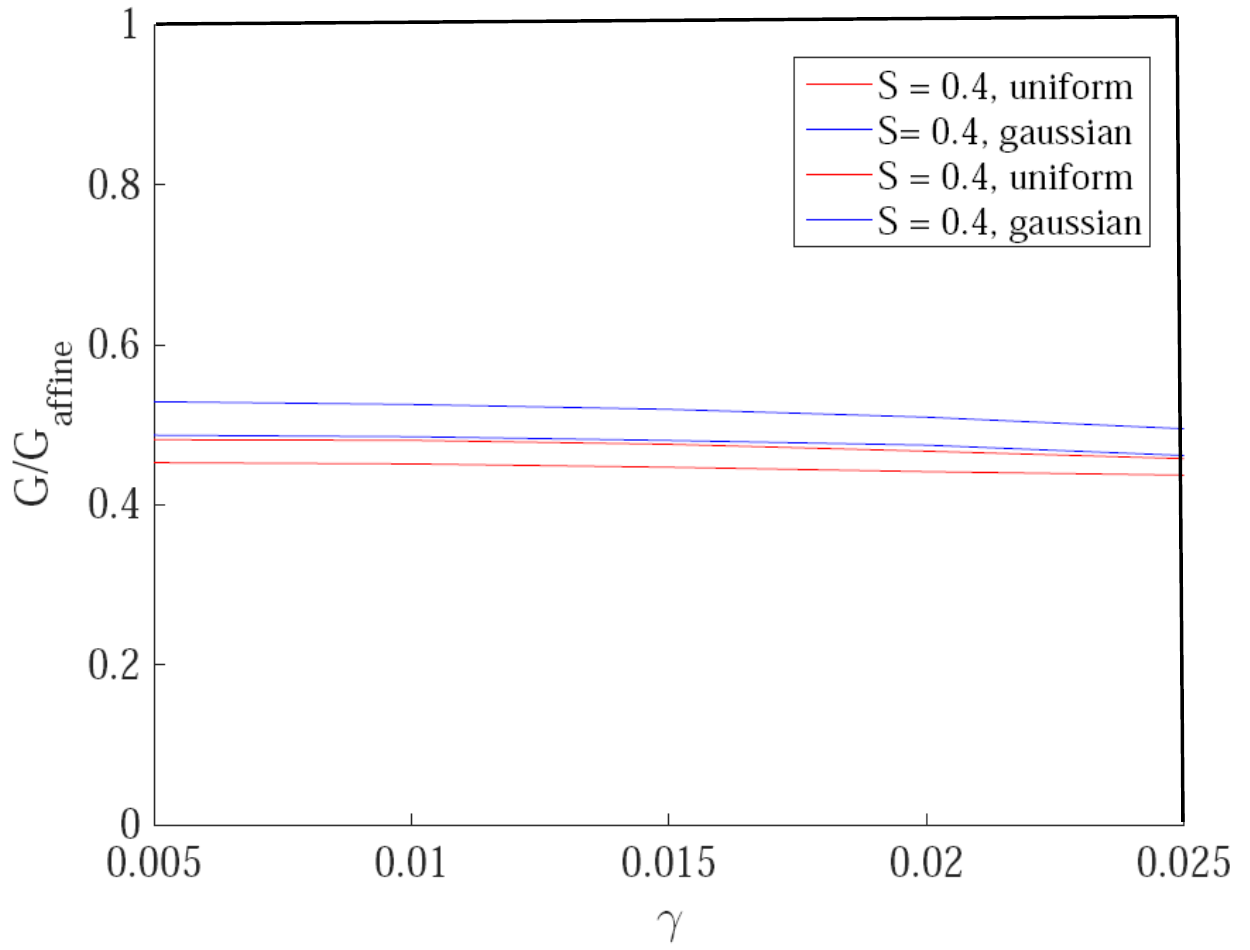} 
	\renewcommand{\thefigure}{S10}
	\caption{\textbf{Linear modulus versus strain for nematic networks prepared with different forms of their orientational order.}  The (red) uniform networks have step orientational distributions so that all angles within $\Delta \theta = \pi/3$ the nematic director are equally probable and all other orientations are disallowed. The (blue) Gaussian distributions ($\sigma^2 = \pi/8$) replace the step distribution with a Gaussian one, as used in all simulations reported in the manuscript.  The comparisons are done for two different samples (one pair of curves for each) to also study the sample to sample variations. The difference in modulus between these networks is small in all cases. 
 }
	\label{FS10}
\end{figure*}

\pagebreak
\pagebreak

\end{document}